\documentclass[useAMS,usenatbib]{mn2e}

\usepackage{epsfig}
\usepackage{myaasmacros}
\usepackage{amsmath}

\def\lesssim{\lower.5ex\hbox{$\; \buildrel < \over \sim \;$}}
\def\gtrsim{\lower.5ex\hbox{$\; \buildrel > \over \sim \;$}}

\title[Semi-analytic models and cosmological simulations] {Galaxy
  formation in semi-analytic models and cosmological hydrodynamic zoom
  simulations} \author[Hirschmann et al.]{Michaela
  Hirschmann$^{1}$\thanks{E-mail: mhirsch@usm.lmu.de}, Thorsten
  Naab$^{2}$, Rachel S. Somerville$^{3,4}$, Andreas Burkert$^{1}$,
  \newauthor Ludwig Oser$^{2}$\\ $^{1}$Universit\"ats Sternwarte
  M\"unchen, Scheinerstr.1, D-81679 M\"unchen, Germany\\ $^{2}$Max
  Planck Institut f\"ur Astrophysik, Karl-Schwarzschild-Str. 1,
  D-85741 Garching, Germany\\ $^{3}$Space Telescope Science Insitute,
  3700 San Martin Dr., Baltimore, MD 21218, USA\\ $^{4}$Department of
  Physics and Astronomy, Johns Hopkins University, Baltimore, MD
  21218, USA}

\begin{document}

\date{Accepted by MNRAS, 10/05/2011}

\pagerange{\pageref{firstpage}--\pageref{lastpage}} \pubyear{2002}

\maketitle

\label{firstpage}

\begin{abstract}
We present a detailed comparison between numerical cosmological
hydrodynamic zoom simulations and 
the semi-analytic model of \citet{Somerville08}, run
within merger trees extracted from the simulations. The high-resolution simulations
represent 48 individual halos with virial masses in the range $2.4\ \times
10^{11} M_{\odot} < M_{\mathrm{Halo}} < 3.3\ \times 10^{13} M_{\odot}$. They include radiative H
\& He cooling, photo-ionization, star formation and thermal SN
feedback. We compare with different SAM versions including
only this complement of physical processes, and also ones including
supernova driven winds, metal cooling, and feedback from Active
Galactic Nuclei (AGN). Our analysis is focused on the cosmic evolution
of the baryon content in galaxies and its division into various
components (stars, cold gas, and hot gas), as well as how those
galaxies acquired their gas and their stellar mass. Both the SAMs and
simulations are compared with observational relations between
halo mass and stellar mass, and between stellar mass and star
formation rate, at low and high redshift. We find some points of
agreement and some important disagreements. SAMs that include the same
physical processes as the simulations reproduce the total baryon
fraction in halos and the fraction of cold gas plus stars in the
central galaxy to better than 20\%.
However, the simulations turn out to have much higher star formation
efficiencies (by about a factor of ten) than the SAMs, despite
nominally being both normalized to the same empirical Kennicutt
relation at $z=0$. Therefore the cold gas is consumed much more
rapidly in the simulations and stars form much earlier. Also,
simulations show a transition between stellar mass growth that is
dominated by in situ formation of stars to growth that is
predominantly through accretion of stars formed in external
galaxies. In SAMs, stellar growth is always dominated by in situ star
formation, because they significantly underpredict the fraction of
mass growth from accreted stars relative to the simulations. In
addition, SAMs overpredict the overall gas accretion rates relative to
the simulations, and overestimate the fraction of ``hot'' relative to
``cold'' accretion. We discuss the reasons for these discrepancies,
and identify several physical processes that are missing in our SAM
and in other semi-analytic models
but which should be included. We also highlight physical processes
that are neglected in the simulations studied here, but which appear
to be crucial in order to understand the properties of real galaxies.
\end{abstract}

\begin{keywords}
galaxies: formation; galaxies: evolution; methods:
N-body-simulations; methods: numerical
\end{keywords}

\section{Introduction}
\label{intro}
The dark energy dominated dark matter paradigm ($\Lambda$CDM) provides
a successful theoretical model for understanding and simulating galaxy
formation. Within this framework the large-scale structure of dark
matter develops from initially small-scale density fluctuations, in a
bottom-up scenario only driven by gravitational forces
\citep{Blumen84}. The baryonic component follows the dark matter, is
shock-heated, cools and condenses into galaxies (\citealp{White78,
  Rees77}).  

The assembly of dark matter halos, which dominate the total matter
content in the Universe, in large cosmological volumes can be followed
with merger trees based on analytic approaches, e.g. using Monte-Carlo
methods based on the extended Press-Schechter formalism (EPS,
\citealp{Press74,Bower91,Bond91,Somerville99,Neistein08,Zhang08,2010MNRAS.401.1796A}).
Alternatively, the full dynamical evolution of dark matter can be
accurately followed with collisionless particles in direct numerical
simulations which are, by now, well resolved at the relevant scales
\citep{Frenk88,Navarro97,Moore99,Klypin99,2003ApJS..145....1B,Springel05,2008MNRAS.391.1685S,2008Natur.454..735D,2010arXiv1002.3660K}. Here
the identification of dark matter halos and the construction of merger
trees is considerably more demanding and various different approaches
have been discussed
(e.g. \citealp{Davis85,Kauffmann93,2000ApJ...544..616G,Springel01,2005MNRAS.364..823W,Genel08,Fakhouri08,2010A&A...519A..94P,2010ApJS..191...43S}
and references therein).  In particular, \citet{Springel05}
constructed and analysed merger trees for the Millennium simulation,
using the ``Friends of Friends'' (\textsc{FOF}) technique
(\citealp{Davis85}) to identify halos, and \textsc{Subfind}
(\citealp{Springel01}) to identify sub-halos (bound objects within
larger virialized dark matter halos).

Simulations of the formation and evolution of the galaxies which are
believed to inhabit these dark matter halos are more demanding,
theoretically as well as numerically. Additional gas-dynamical and
radiative processes, such as the formation of stars and black holes as
well as the respective feedback, have to be taken into account. To
follow the evolution of galaxies two main approaches have been
developed over the past decades: \textit{Semi-analytic models (SAMs)}
and \textit{direct cosmological simulations}. SAMs (\citealp{White91,
  Kauffmann93, Cole94,Kauffmann96,Somerville99a, Kauffmann99,
  Kauffmann00, Cole00, Springel01,
  Hatton03, Kang05, Baugh05, 2006ApJ...648L..21K,Croton06,Bower06,
  DeLucia07, Somerville08, Font08, Guo09, Weinmann09}) use
pre-calculated dark matter merger trees either from EPS or direct
cosmological simulations and follow the formation of galaxies with
simplified, physically and observationally motivated, analytic
recipes. The computational costs of this approach are typically low,
and the influence of different physical mechanisms can be investigated
separately in a straightforward way. Modern SAMs are quite successful
at reproducing observed statistical properties of galaxies in large
cosmological volumes over a large range of galaxy masses and redshifts
(e.g. \citealp{Somerville08,Guo10}). Disadvantages are that the
dynamics of the baryonic component (gas and stars) and the interaction
between baryonic matter and dark matter are not followed directly and
that in many cases the assumed models are simplified and use a large
number of free parameters to fit different observations simultaneously
(\citealp{Somerville08,Benson10,Bower10}).

Direct cosmological galaxy formation simulations can follow the
evolution of dark matter and gas explicitly. Even though they treat
the underlying dynamics more correctly than SAMs, the spatial and mass
resolution, at present, is not high enough to accurately simulate
intermediate and low mass galaxies in large cosmological volumes. In
addition, small scale processes like e.g. the formation of stars and
black holes with the associated feedback has to be computed in a
simplified manner with sub-grid/sub-resolution models
\citep{Cen93,Dave01,Springel03,Maller04,2005ApJ...627..608N,2005MNRAS.363....2K,Navarro09,2010MNRAS.402.1536S},
which again require the introduction of parameters.  Ab initio
cosmological zoom simulations with proper cosmological boundary
conditions enable direct simulations of the baryonic physics of
certain regions of interest at higher resolution, either limited to
small cosmological volumes \citep{2009MNRAS.399.1773C} or, more
popularly, individual halos
\citep{1997ApJ...478...13N,2007MNRAS.374.1479G,2007ApJ...658..710N,
  2009ApJ...694..396B,Oser10,Wadepuhl11,2010MNRAS.406..936P,2010arXiv1003.4744T,
  Sawala10,Piontek11, Agertz11}. These simulations can
attain very high resolution, and provide a way to resolve galaxies of
very different masses with the appropriate resolution in each
case. However they are very time consuming and therefore not currently
feasible for representative studies of large populations of
galaxies. In addition, the sub-resolution models are uncertain and it
is still unclear how sensitive various results may be to the details
of these sub-grid models or the parameter values.

Both approaches make definite predictions for the evolution of galaxy
properties at various masses over cosmic time. Because of their
greater computational efficiency, SAMs generally include more models
for physical
processes than current numerical simulations, and because of their
greater flexibility, it has been possible to tune them to obtain quite
good agreement with a broad range of galaxy properties in the local
Universe. SAMs have also been shown to reproduce the statistical
properties (e.g. luminosity and stellar mass functions, star formation
rates) of high redshift galaxies ($z\lesssim6$) quite well, at least
for massive galaxies ($m_{\rm star} \gtrsim 10^{10} M_{\odot}$;
e.g. \citealt{somerville:11, fontanot:09}). Therefore we might expect
SAMs to do a better job of reproducing the observed universe than the
simulations, but we might worry that they could do so for the wrong
reasons. Because there is a great deal of uncertainty in many of the
important processes, and most of the physical recipes contain free
parameters, if one physical process (e.g. gas cooling and accretion)
is modelled inaccurately in the SAM, it is currently possible to
compensate by tuning a competing process (such as feedback). By
running SAMs within merger trees extracted from numerical hydrodynamic
simulations, in order to constrain the evolution of the dark matter
component to be the same in both cases, we can isolate various
physical processes and attempt to improve the accuracy of the
semi-analytic recipes. At the same time, by comparing the detailed
predictions of the formation histories of galaxies in the simulations
with the SAM predictions, we may gain insights into the origin of
existing discrepancies between the simulations and the real universe.

Various comparison studies between simulations and SAMs for large
galaxy populations as well as individual halos have been discussed in
the literature (we summarize these results in section \ref{previous})
following different philosophies. For some studies only individual
physical processes, like cooling, were investigated (\citealp{Lu10,
  Benson10}), while others focussed on the evolution of individual
objects, as a high-mass galaxy cluster (\citealp{Saro10}) or a single disk
galaxy (\citealp{Stringer10}).

Our approach is new in many respects. We compare the evolution of
individual halos but use the, up to now, largest number of
high-resolution zoom simulations ($48$), presented in
\citet{Oser10}. The simulations cover dark matter halos in the mass
range of $2.4\ \times 10^{11} M_{\odot} < M_{\mathrm{Halo}} <
3.3\ \times 10^{13} M_{\odot}$. Although a limited complement of
physical processes have been taken into account in the simulations,
the more massive of these halos have been shown to represent fairly
well the evolution of observed massive galaxies \citep{Oser10}. These
simulations are compared to results from the full SAM of
    \citet{Somerville08}, which has been shown to represent present day
galaxy properties reasonably well over a wide range of masses, as well
as different stripped down versions. The SAMs are run within merger
trees extracted directly from the numerical simulations. In addition,
we compare both model predictions to observations at different
redshifts and point out where the respective models succeed or fail
either to match each other and/or the observations.

The paper is organized as follows: In section \ref{previous} we
discuss results from previous comparisons between SAMs and
simulations. The hydrodynamical simulations and the merger-tree
construction method used for this study are discussed in section
\ref{sim} and we briefly review the ingredients of the
\citet{Somerville08} SAM in section \ref{sam}. The redshift evolution
of the baryonic components in simulations and SAMs is compared in
section \ref{galprop} followed by a comparison to observations in
section \ref{comp}. In section \ref{conclusion} we summarize and
discuss our main results. A resolution study for individual halos can
be found in the Appendix.

\section{Previous comparison studies}
\label{previous}

Previous quantitative comparisons between simulations and SAMs have
either focused on whole populations of galaxies \citep{Benson01,
  Yoshida02, Helly03, Cattaneo07,Benson10, Lu10} or individual objects
like a galaxy cluster \citep{Saro10} and a single disk galaxy
\citep{Stringer10}. \citet{Helly03} compared the efficiency of gas
cooling for different dark matter halos as a function of redshift
between a $(50 \mathrm{Mpc}/h)^3$ SPH
(Smoothed-Particle-Hydrodynamics) simulation (\textsc{Hydra},
\citealp{Pearce01}) excluding star formation, heating and feedback and
a stripped down version -- without star formation or feedback -- of
the \textsc{Galform} SAM \citep{Cole00}. For $z=0$ they find good
agreement of the cold gas mass between the SPH simulation and the
SAM. At high redshifts, however, more gas tends to cool in low-mass
halos in the simulation due to the limited numerical
resolution. Still, they conclude that simulations and SAMs give
consistent results for the evolution of cooling galactic gas and
confirm earlier findings of \citet{Benson01} and \citet{Yoshida02}.

\citet{Cattaneo07} did consider star formation and supernova feedback
for a similar comparison in a $34.19 \mathrm{Mpc}^3$ volume. For the
SPH-simulation they used TreeSPH (\citealp{Dave97} and the input for
their SAM-code GalICS (\citealp{Hatton03}) was the merger trees
constructed from the dark matter component of the SPH-simulation. The
SAM did not include a photo-ionising background but followed the
cooling by metals, while the simulations did include a photo-ionizing
background and assumed primordial composition of He and H. The star formation
prescription in the SAM was quite standard (star formation occurs
above a certain gas surface density, according to a
Kennicutt-Schmidt-like relation), and a simple recipe for SN-driven
winds was also included. In order to replicate AGN feedback, star
formation is quenched when the bulge component of a galaxy reaches a
critical mass. For the comparison they used two different SAM
versions: one with no feedback and the `full' model. In general, they
found good agreement between simulations and the no-feedback model for
the baryonic mass functions at different redshifts and in different
environments. Moreover, simulations and the no-feedback SAM made
similar predictions for the `hot' and `cold' mode gas accretion
histories of galaxies (e.g. \citealp{Keres05}). However, at low
redshifts, much less gas was left over in the simulation than in the
no-feedback SAM with both approaches over-predicting the observed
baryonic mass function, in particular at the high mass end. The full
SAM, on the other hand, matched the observations due the inclusion of
supernova-driven outflows and AGN feedback, which suppresses gas
cooling in large halos. They concluded that the simulations and the
no-feedback model failed as a consequence of missing physics rather
than computational inaccuracies.

\citet{Saro10} compared the galaxy populations within a massive
cluster ($M_{\mathrm{cluster}} = 1.14 \times 10^{15} M_{\odot}$) using
a high-resolution cosmological re-simulation run with \textsc{Gadget}2
(\citealp{Dolag09}) and the SAM model of \citet{DeLucia07}. They
focused on differences between the central and the satellite galaxies
considering only gas cooling and star formation and neglecting any
form of feedback. In general, they find similar \textit{statistical
  properties} for the galaxy populations, e.g. the stellar mass
function with a few remarkable \textit{object by object} differences.
The central galaxy in the simulation starts with a more intense and
shorter initial burst of star formation at high redshift and forms
fewer stars at low redshift than in the SAM. While in the SAM all
stars in the central galaxy are formed in its progenitors, in the
simulations the final stellar mass is larger than the sum of all
progenitors. Satellite galaxies can lose up to $90$ per cent of their
stellar mass due to tidal stripping -- a process, which is, however,
not included in the \citet{DeLucia07} semi-analytic model, nor in most
models discussed in the recent literature.

Moreover, \citet{Stringer10} presented a comparison for the evolution
of a single disk galaxy using the SPH-code \textsc{Gasoline}
(\citealp{Wadsley04}) and the semi-analytic model \textsc{Galform}
(\citealp{Bower06}) based on the dark matter merger history of the
simulation. They find that the two techniques show a
\textit{potential} consistency for the evolution of the stellar and
gas components by assuming the same physics and the same initial
conditions. They try to mimic in the SAM the `blast wave' SN feedback
implemented in the simulation, i.e. after a supernova explosion no
cooling is allowed in a certain volume. However, using the
\textsc{Galform} model as described in \citet{Bower06} (including
chemical enrichment, supernova and AGN feedback), the resulting system
is not recognisably the same as predicted by simulations. At all
redshifts, the stellar mass is much larger and the hot gas fraction is
much lower in the simulation than in the SAM.

Finally, \citet{Lu10} and \citet{Benson10} focus on the algorithms for
gas cooling in SAMs in great detail. \citet{Benson10} compare cold
(rapid) and hot (slow) accretion rates in the \textsc{Galform} SAM and
in simulations from \citet{Keres09} ($50\ \mathrm{Mpc}/h, 2 \times
288^3$ particles). They used their `full' model including feedback and
metal cooling, although these processes are not included in the
simulations. Moreover, they modified their SAM by adopting an updated
calibration for the transition between the rapid and slow cooling
regime following the methodology of \citet{Birnboim03}. They find
reasonably good agreement for the hot and cold mode accretion fraction
in the SAM and the simulations and thus, they conclude that the
cold-mode physics is already adequately accounted for in SAMs.  In the
study of \citet{Lu10} five different SAMs (`Munich' model:
\citealp{Croton06}, `Kang' model: \citealp{Kang05}, `\textsc{Galform}'
model: \citealp{Cole00}, `GalICS' model: \citealp{Hatton03} and the
`Somerville' model: \citealp{Somerville99a}) are compared to the
simulations of \citet{Keres09}, without considering any feedback or
metal enrichment in either method. They find a significant difference
between hot and cold accretion rates: compared to the simulations, the
cold mode accretion rates are too low and the hot mode accretion are
too high in SAMs. They construct a modified cooling recipe for the SAM
to enable simultaneous hot and cold accretion, resulting in much
better agreement between the SAMs and the simulations.

Throughout the course of this paper, we will refer back to these
studies and comment upon the similarities and differences with our
results.

\section{The simulation and merger tree construction}\label{sim} 

\subsection{Simulation setup}\label{setup}

The cosmological zoom simulations presented in this paper are
described in detail in \citet{Oser10} and we briefly review the
simulation setup here. The dark matter halos for further refinement
were selected from a dark matter only N-body simulation
(\textsc{Gadget}-2, \citealp{Springel05}) with a comoving periodic
box length of $L=100\ \mathrm{Mpc}$ and $512^3$ particles (see also
\citealp{Moster10}). We assume a $\Lambda$CDM cosmology based on the
WMAP3 measurements (see e.g. \citealp{Spergel03}) with $\sigma_8 =
0.77$, $\Omega_{m}=0.26$, $\Omega_{\Lambda}=0.74$, and
$h=H_0/(100\ \mathrm{kms}^{-1})=0.72$. The simulation was started at
$z=43$ and run to $z=0$ with a fixed comoving softening length of
$2.52\ h^{-1} \mathrm{kpc}$ and a dark matter particle mass of
$M_{\mathrm{DM}} = 2 \times 10^8 M_{\odot}/h$. Starting at an
expansion factor of $a=0.06$ we constructed halo catalogues for 94
snapshots until $z=0$ separated by $\Delta a =0.01$ in time. From this
simulation, we picked $48$ halos identified with the halo finder
algorithm $FOF$ at $z=0$. To construct the high-resolution initial
conditions for the re-simulations, we trace back in time all particles
that are closer than $2 \cdot r_{200}$ to the center of the halo in
any snapshot and replace them with dark matter as well as gas particles at
higher resolution ($\Omega_b=0.044, \Omega_{DM}=0.216$). In the high
resolution region the dark matter particles have a mass resolution of
$m_{\mathrm{DM}} = 2.1\cdot 10^7 M_{\odot}h^{-1}$, which is 8 times
higher than in the original simulation, and the gas particle masses
are $m_{\mathrm{Gas}} = m_{\mathrm{Star}} = 4.2\cdot 10^6
M_{\odot}h^{-1}$. Individual cases were run at 64 times higher mass
resolution and 4 times higher spatial resolution. The re-simulated
halos cover a mass range of two orders of magnitude ($2.4 \times 10^{11}
M_{\odot} < M_{\mathrm{Halo}} < 3.3 \times 10^{13} M_{\odot}$).

For modeling the gas component we use the entropy conserving
formulation of SPH (\textsc{Gadget}-2, \citealp{Springel05}). We
include star formation and cooling for a primordial composition of
hydrogen and helium (\citealp{Theuns98}). The cooling rates are
computed under the assumption that the gas is optically thin and in
ionization equilibrium. Furthermore, our simulations include a
spatially uniform redshift dependent UV background radiation field
according to \citet{Haardt96}, where reionization takes place at $z
\approx 6$ and the radiation field peaks at $z \approx 2-3$.

\begin{figure*}
\begin{center}
  \epsfig{file= 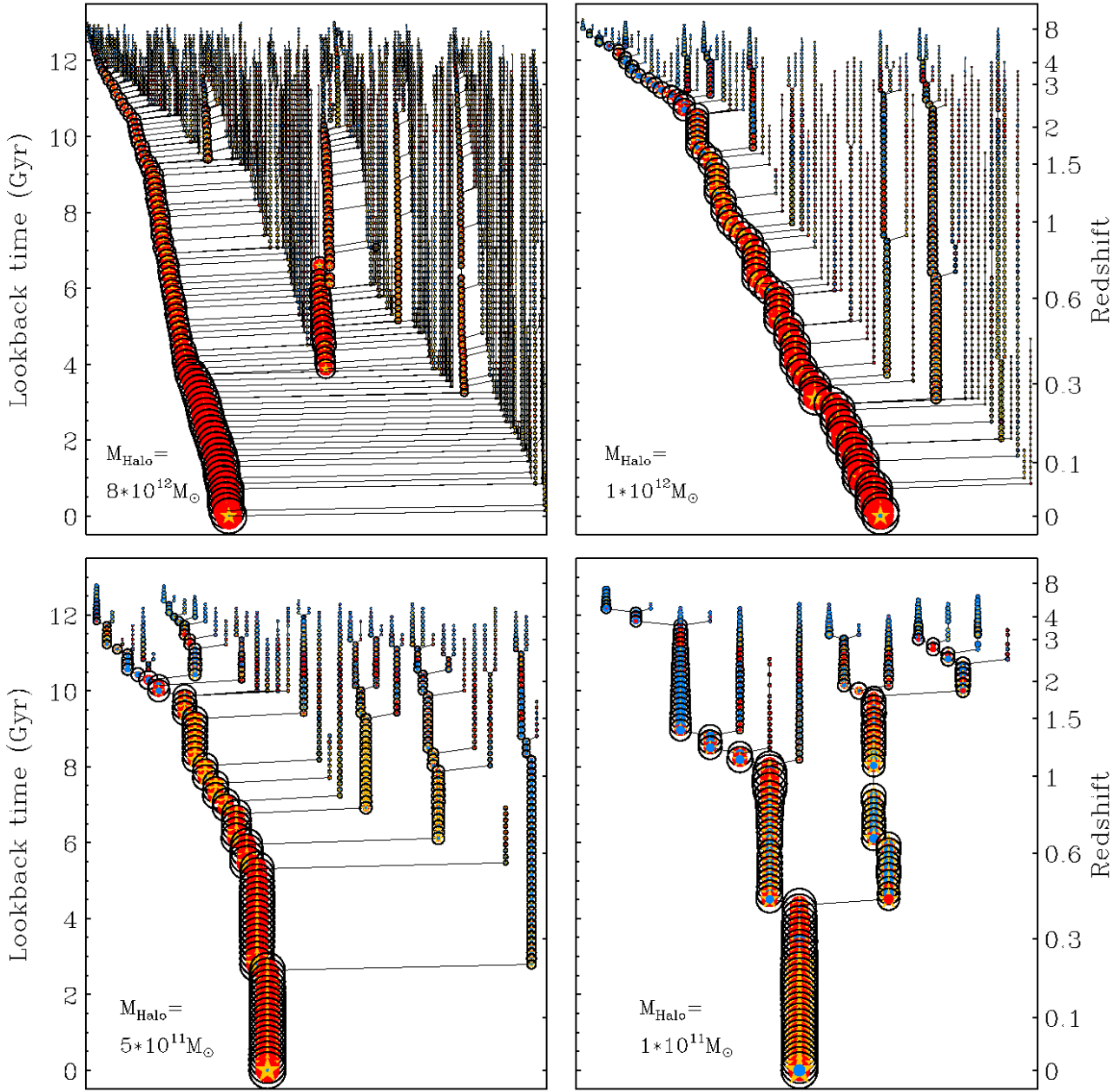, width=1.0\textwidth}
  \caption{Visualisation of merger trees for four re-simulated halos
    with different masses: upper left:$M_{\mathrm{vir}} = 8 \times
    10^{12} M_{\odot}$, upper right: $M_{\mathrm{vir}}= 1 \times
    10^{12} M_{\odot}$, lower left: $M_{\mathrm{vir}}= 5 \times
    10^{11} M_{\odot}$, lower right: $M_{\mathrm{vir}} = 1 \times
    10^{11} M_{\odot}$. Black circles show the dark matter halo at
    every time-step of the simulations. The symbol size is
    proportional to the square root of the halo mass normalized to the
    halo mass at z=0.  The yellow stars indicate the stellar mass, the
    blue and red filled circles the cold and hot gas mass within the
    virial radius of the dark matter halo. The symbol sizes for the
    baryons scale with the square root of the masses normalized to the
    maximum total baryonic mass at z=0.}  {\label{TreeVisGroup}}
\end{center}
\end{figure*}

To model star formation and SN feedback we use the approach of
\citet{Springel03}.  In this model, the ISM is treated as a two-phase
medium where clouds of cold gas form from cooling of hot gas and are
embedded in the hot gas phase assuming pressure equilibrium. The hot
gas is heated by supernovae and can evaporate the cold clouds.  Stars
form from the cold gas whenever the local density exceeds a threshold
density ($\rho > \rho_{\mathrm{th}} = 0.205\mathrm{cm}^{-3}$). The
star formation rate is calculated by
\begin{equation}
\frac{d \rho_*}{dt} = (1-\beta)\frac{\rho_c}{t_*}
\end{equation}
Here, $\beta$ is the mass fraction of massive stars, which are assumed
to explode as supernovae type II, $\rho_c$ is the density of cold gas
and $t_* = t^0_*(\rho/\rho_{\mathrm{th}})^{-1/2}$ is the star
formation time scale. The supernova explosions heat the surrounding
gas with an energy input of
$10^{51}\ \mathrm{ergs}$. \citet{Springel03} used an idealized,
isolated disk galaxy simulation to set the free parameters
$\rho_{\mathrm{th}}\ \mathrm{and}\ t^0_*$, by adjusting them to obtain
a match to the observed Schmidt-Kennicutt relation.  We adopt the same
values of these parameters here.

\subsection{Merger trees}\label{trees}

We extract the merger trees for the dark matter component directly
from the cosmological re-simulations as described in
\citet{Hirschmann10}. For every snapshot at a given 
redshift, we first identify individual dark matter haloes using a FOF
(Friends-of-Friends) algorithm with a linking length of $b=0.2$ ($
\approx 28 \mathrm{kpc}$, \citealp{Davis85}). In a second step we
extract the subhalos of every FOF group using the \textsc{Subfind}
algorithm \citep{Springel01GAD}. This halofinder identifies over-dense
regions and removes gravitationally unbound particles. In this way we
split the FOF group into a main or host halo and its satellite
halos. In most cases, $90\%$ of the total mass is located in
the main halo.

The sizes and virial masses of the main halos (i.e. the most massive
\textsc{Subfind} halos) are determined by a spherical overdensity
criterion. The minimum halo mass is set to 20 particles ($5 \times
10^8 M_{\odot}/h$). In the following, we will use \textit{isolated}
merger trees which are constructed only for the the main halos,
i.e. the central objects of one FOF group identified by
\textsc{Subfind}. The mass of a central object is defined by the dark
matter mass within the virial radius using the overdensity
approximation in the spherical collapse model according to
\citet{Bryan98}. The algorithm to connect the dark matter halos
between the snapshots at different redshifts is described in detail in
\citet{Maulbetsch07}. The branches of the trees for $z=0$ halos are   
constructed by connecting the halos to their most massive progenitors (MMP) 
at previous snapshots. Thereby, halo $j$ with $n_j$ particles at
redshift $z_j$ with the maximum probability $p(i,j)$  
is chosen to be a MMP of halo $i$ containing $n_i$ particles at
redshift $z_i$ (where $j < i$). The probablity $p(i,j)$ is defined as  
\begin{align}
p(i,j) & = \frac{n_{ov}(i,j)}{n_{max}(i,j)} \quad \text{with} \\
n_{ov} & = n_i(z_i) \cap n_j(z_j) \quad \text{and} \nonumber\\
n_{max}(i,j) & = \max(n_i(z_i),n_j(z_j)) \nonumber
\end{align}
Here, $n_{ov}$ is the number of particles found in both halos and
$n_{max}$ is the particle number of the larger halo. We remove `fake'
haloes which exist only within one timestep and have no connection to
any branch (halo masses are generally near to the resolution
limit). The low redshift ends of the branches are then checked for mergers. 
A halo $j$ at $t_j$ is assumed to merge into halo $i$ at $t_i$, if at
least $50 \% $ of the particles of halo $j$ are found in halo $i$. In
case of a merger the branches are connected.  

Note that the tree-algorithm is only applied to the dark matter
particles -- star or gas particles are not separately traced back in
time. They are assumed to follow the evolution of the dark
matter. Therefore, we assign to each dark matter halo in a tree a
hot/cold phase gas mass by counting hot/cold gas particles within the
virial radius of the central halo. The stellar and cold
gas particles within $1/10$ of the virial radius are defined as the
stellar and gas mass of the central galaxy. We distinguish between hot
and cold gas particles by using the following definition:
\begin{eqnarray}\label{coldhot}
\log T < 0.3 \log\rho + 3.2\ \rightarrow \mathrm{cold}\\
\log T > 0.3 \log\rho +3.2\ \rightarrow \mathrm{hot}
\end{eqnarray}
 Note that the above discrimination between hot and
  cold gas was established by looking directly at the phase diagrams
  of the re-simulations, where we have divided between the gas in the
  disk heated by SN feedback and the shock heated gas. With the above
  definition for cold gas we mainly capture the dense, star-forming
  gas.

In Fig. \ref{TreeVisGroup} we show a visualization of four merger
trees of re-simulated halos with virial masses of $8 \times
10^{12}M_{\odot}$, $1 \times 10^{12}M_{\odot}$, $5 \times
10^{11}M_{\odot}$ and $1 \times 10^{11}M_{\odot}$.  The size of the
black circles approximates the dark matter halo mass, the yellow stars
the stellar mass within the virial radius and the blue and red filled
circles the cold and hot gas component, respectively. The symbol sizes
scale with the square root of mass normalized to the final dark matter
halo mass (dark matter component) and to the final baryonic mass
(star, hot and cold gas mass). We clearly see that galaxies at high
redshift contain more cold gas, which either turns into stars or is
heated towards lower redshifts. In general, for more massive halos the
fraction of cold gas and stars at $z=0$ is lower.

To study the influence of numerical resolution on the evolution of the
dark matter and the baryonic components we have simulated a few halos
with $4\ \times$ higher spatial resolution ($=64\ \times$ higher mass
resolution) than the original dark matter simulation. A comparison of
the results can be found in the Appendix. The overall mass assembly of
the main halos and the number of major mergers do not show any
significant variation, although the number of identified minor mergers
increases due to the higher resolution. Overall, we conclude that our
results are well-converged and would not change significantly if we
improved the resolution.

\section{The semi-analytic model}\label{sam} 

The merger-trees constructed as described above are used as input for
the semi-analytic model described in \citet[][hereafter
  S08]{Somerville08}. The SAM makes use of merger trees for
``isolated'' halos only, and treats the evolution of sub-structure
within virialized halos using semi-analytic approximations. The `full'
version includes photo-ioniziation, gas cooling, star formation, SN
feedback, metal enrichment, and black hole growth in a radio and
quasar mode with corresponding feedback. However, to provide a more
meaningful comparison to our simulations, we do not only consider the
`full' version, but also `stripped down' models by separately
switching off AGN feedback, metal cooling, Supernova-driven winds, and
`thermal' Supernova feedback. We consider the following different
versions:

\begin{itemize}
\item{{\bf{NF}}: no Feedback, primordial metallicity}
\item{{\bf{SN}}: thermal SN-feedback, primordial metallicity}
\item{{\bf{SNWM}}: thermal SN-feedback, SN-driven Winds, metal cooling}
\item{{\bf{FULL}}: `full' version, including thermal SN-feedback, SN
  winds, metal cooling, and AGN feedback}
\end{itemize}

In the following we briefly summarize how the different physical
mechanisms are implemented and how they differ from the ones in
    the simulations. For full details we refer the reader to 
\citet{Somerville08}. In Table \ref{param}, we provide a summary of
the galaxy formation parameters used here. In Table
    \ref{directcomp}, we give an explicit overview of the physical
    recipes assumed in the different SAM versions starting from the NF
    model. These are compared to the physics which are implemented in
    simulations.

\begin{table*}
\centering
\caption{Summary of the galaxy formation parameters in the fiducial
  model, which are partly deviating from the ones in S08}

\begin{tabular}{ p{5cm} p{7cm}p{4cm}}\hline\\\bf{Parameter} & \bf{Description} &
  \bf{Fiducial value} \\\hline \hline
\textit{Quiescent star
      formation}\\
$A_{\mathrm{KS}}$ & Normalization of Kennicutt law
  & $1.67 \times
  10^{-4}\ M_{\odot}\ \mathrm{yr}^{-1}\ \mathrm{kpc}^{-2}$\\
$N_K$ & Power-law index in Kennicutt law &1.4\\
$\Sigma_{\mathrm{crit}}$ & Critical surface density &
  6$M_{\odot}\ \mathrm{pc}^{-2}$ \\ 
\hline\textit{Burst star formation}\\
$\mu_{\mathrm{crit}}$ & Critical mass ratio for burst activity &
0.1\\ 
\hline\textit{SN feedback}\\
$\epsilon_{\mathrm{SN}}^{0}$ & Normalization of
    reheating fct & 1.3\\
$\alpha_{\mathrm{rh}}$ & Power-law slope of
    reheating fct & 2.0\\
$V_{\mathrm{eject}}$ & Velocity scale for
    ejecting gas &
    120
$\mathrm{km}\ \mathrm{s}^{-1}$\\
$\chi_{\mathrm{re-infall}}$
    & Time-scale for re-infall of ejected gas
&
    0.1\\ \hline
\textit{Chemical evolution}\\
$y$ & Chemical yield
    & 1.5 \\ \hline
\textit{Black hole
      growth}\\
$\eta_{\mathrm{rad}}$ & Efficiency of conversion of
    rest mass to
radiation & 0.1\\
$M_{\mathrm{seed}}$ & Mass of
    seed black hole & 100 $M_{\odot}$\\
$f_{\mathrm{BH,final}}$ &
    Scaling factor for mass after merger &
    0.8\\
$f_{\mathrm{BH,crit}}$ & Scaling factor for critical BH
    mass & 0.4\\  \hline
\textit{AGN-driven
      winds}\\
$\epsilon_{\mathrm{wind}}$ & Coupling factor for
    AGN-driven winds &
0.5 \\ \hline
\textit{Radio-mode
      feedback}\\
$\kappa_{\mathrm{radio}}$ & Normalization of 'radio
    mode' accretion
rate & $2 \times
    10^{-3}$\\
$\kappa_{\mathrm{heat}}$ & Coupling efficiency of
    radio jets with hot
gas & 1.0
    \\
\hline
\end{tabular}
\label{param}
\end{table*}

\begin{table*}
\centering
\caption{Overview of the different physical mechanisms for galaxy
  formation assumed in the different SAM versions and implemented in
  the simulations.}
\begin{tabular}{p{1cm} | 
    p{2cm}p{2.5cm}p{1.5cm}p{3cm}p{1.5cm}p{2.5cm}}\hline\\\bf{Model} & \bf{Gas cooling} &
  \bf{Star formation} & \bf{Metals} &
  \bf{Thermal SN feedack} & \bf{SN winds} & \bf{AGN feedback} \\\hline \hline
\bf{NF} & Yes & Yes & No & No & No & No\\
\hline
\bf{SN} & Yes & Yes & No & Yes & No & No\\
\hline
\bf{SNWM} & Yes & Yes & Yes & Yes & Yes & No\\
\hline
\bf{FULL} & Yes & Yes & Yes & Yes & Yes & Yes\\
\hline
\bf{SIM} & Yes & Yes & No & Yes & No & No\\
\hline
\end{tabular}
\label{directcomp}
\end{table*}

\begin{enumerate}

\item{{\bf{Radiative cooling}}: 
The rate of gas condensation via atomic cooling is computed based on
the model proposed by \citet{White91}. The cooling time is computed as
\begin{equation}\label{tcool}
t_{\mathrm{cool}} = \frac{3/2 \mu m_p kT}{\rho_g(r) \Lambda(T,Z_h)}.
\end{equation}
Here, $T$ is the virial temperature, $\mu m_p$ is the mean molecular
mass, $\rho_g(r)$ is the radial density profile of the gas and
$\Lambda(T,Z_h)$ is the cooling function, which is temperature and
metallicity dependent. The cooling time is the time required for the
gas to radiate away all its energy starting at the virial
temperature. The gas density profile $\rho_g(r)$ is assumed to follow
an isothermal sphere: $\rho_g(r) = m_{\mathrm{hot}}/(4\pi r_{vir}
r^2)$. Putting this expression in Eq. \ref{tcool} one can solve for a
cooling radius $r_{\mathrm{cool}}$. Within the cooling radius all gas
can cool within the cooling time $t_{cool}$. The cooling rate for the
mass within $r_{\mathrm{cool}}$ is
\begin{equation}\label{coolrate}
\frac{dm_{\mathrm{cool}}}{dt} = \frac{1}{2} m_{\mathrm{hot}} \frac{r_{\mathrm{cool}}}{r_{\mathrm{vir}}} \frac{1}{t_{\mathrm{cool}}}.
\end{equation}

Following \citet{Springel01} and \citet{Croton06} it is assumed that
the cooling time is equal to the halo dynamical time
$t_{\mathrm{cool}} = t_{\mathrm{dyn}} =
r_{\mathrm{vir}}/V_{\mathrm{vir}}$.  Two different modes of accretion
are distinguished: the rapid (``cold mode'') and the slow (``hot
mode'') cooling regime. In the rapid cooling regime, where the cooling
radius is larger than the virial radius $r_{\mathrm{cool}} >
r_{\mathrm{vir}}$, the cooling rate is set to the gas accretion rate,
which is governed by the mass accretion history. Slow cooling occurs
whenever the cooling radius is smaller than the virial radius
$r_{\mathrm{cool}} < r_{\mathrm{vir}}$. Here, the cooling rate is
calculated according to eq. \ref{coolrate}.  The same cooling
    function is used in the simulations, however, the cooling rate is
    calculated locally based on the density and temperature.}

\item{{\bf{Photo-ionization}}: 
Photo-ionization heating is considered
  in all four SAM versions.  It causes halos below a certain filtering
  mass $M_F$ to have a lower baryon fraction than the universal
  average. The collapsed baryon fraction as a function of redshift and
  halo mass is parameterized by the expression:
\begin{equation}
f_{\mathrm{b,coll}}(z,M_{\mathrm{vir}}) = \frac{f_b}{[ 1 + 0.26 M_F(z)/M_{\mathrm{vir}} ]^3},
\end{equation}
where $f_b$ is the universal baryon fraction. The filtering mass is a
function of redshift and depends on the reionization history of the
universe (\citealp{Kravtsov04}).  Note that photo-ionization heating
has very little effect on galaxies with circular velocities larger
than about 30--50 km/s, and therefore plays a minor role in our study,
which mainly focusses on larger galaxies. In the simulations a UV
    heating background is implemented instead of a filtering
    mass. However, the filtering mass treatment adopted in the SAM is
    based on the results of numerical hydrodynamic simulations
    \citep{Kravtsov04}, so we do not expect this to introduce any
    significant discrepancy (see also \citealp{Hambrick09} and
    references therein). }

\item{{\bf{Star formation}}: The SAM distinguishes between quiescent
  star formation in isolated disks and merger-driven starbursts.  The
  quiescent star formation is based on the the empirical
  Schmidt-Kennicutt (SK) relation (\citealp{Kennicutt89,
    Kennicutt98}). The star formation rate density is calculated
  according to
\begin{equation}\label{SK}
\Sigma_{\mathrm{SFR}} = A_{\mathrm{Kenn}} \Sigma_{\mathrm{gas}}^{N_K}
\end{equation} 
with $A_{\mathrm{Kenn}} = 1.67 \times 10^{-4}$, $N_K = 1.4$, and
$\Sigma_{\mathrm{gas}}$ is the surface density of cold gas in the
disk. The normalisation uses the conversion factor appropriate for a
Chabrier IMF (\citealp{Chabrier03}). The gas follows an exponential
disk (proportional to the scale-length of the stellar disk) and only
gas above a critical surface density threshold
$\Sigma_{\mathrm{crit}}$ ($=6M_{\odot}/{\rm pc}^2$) is available for
star formation.

Star formation during starbursts is driven by merger events. The star
formation rate is assumed to be a function of the mass ratio and the
combined cold gas content of the merging galaxies, the bulge to total
stellar component and burst timescale. The starburst efficiency as a
function of these variables is based on hydrodynamic simulations of
binary galaxy mergers (see references in S08). While in the SAMs
    we use a 2D implementation for quiescent star formation following
    the Schmidt-Kennicutt law and a simple recipe for starbursts, in
    simulations, the star formation efficiency (in both quiescent and
    burst modes) is determined by the local 3D cold gas density. The
    normalization of the Schmidt-Kennicutt law was chosen by requiring
    a smooth, isolated disk to lie on the observed relation (see
    \citealp{Springel03}).
}

\item{{\bf{Supernova feedback}}: Exploding supernovae deposit thermal
  energy in the interstellar medium, which may heat the cold gas, and in
  certain situations may drive winds that unbind the gas from the
  potential well of the dark matter halo. In the SAM, these processes
  are modeled by removing cold gas from the galaxy and depositing it
  either in the hot gas reservoir, where it can cool again fairly
  quickly, or ejecting it from the halo, where it can fall back again
  on a longer timescale. We will refer to the former as ``thermal'' SN
  feedback and the latter as ``SN-driven winds''. The SN model
  includes only the thermal SN FB, while the SNWM includes both
  thermal SN feedback and SN-driven winds. While supernova driven
  winds have been implemented in some numerical hydrodynamic
  simulations (e.g. \citealp{Oppenheimer08}), only thermal SN feedback
  is implemented in the simulations used in this study.

The heating rate of the cold gas is given by
\begin{equation}\label{mrh}
\dot{m}_{\mathrm{rh}} = \epsilon^{\mathrm{SN}}_0
\left(\frac{V_{\mathrm{disk}}}{200\,\mathrm{km/s}}
\right)^{\alpha_{\mathrm{rh}}} \dot{m}_*, 
\end{equation}
where $\epsilon^{\mathrm{SN}}_0$ and $\alpha_{\mathrm{rh}}$ are free
parameters and $\dot{m}_*$ is the star formation rate. We assume the
circular velocity $V_{\mathrm{disk}}$ to be the maximum rotation
velocity of the dark matter halo, $V_{\mathrm{max}}$.  To
    estimate reasonable values for the parameter
    $\epsilon^{\mathrm{SN}}_0$ we follow the prescription of
    \citet{Kauffmann93}, in which it is assumed that the energy
    released from supernovae heats the gas to the virial temperature
    of the halo (corresponding to a value of $\alpha_{\mathrm{rh}} =
    -2$).  Using this recipe, the thermal energy rate is given as
    \begin{equation}\label{Etherm}
      \dot{E}_{\mathrm{thermal}} = 
      \epsilon_{\mathrm{thermal}}\ \eta_{\mathrm{SN}}\
      E_{\mathrm{SN}}\ \dot{m}_*
      = \dot{m}_{\mathrm{rh}}\ V_{\mathrm{disk}}^2,
    \end{equation}
where $\eta_{\mathrm{SN}}$ is the number of supernovae expected per
solar mass of stars formed ($= 4\times 10^{-3} M_{\odot}^{-1}$),
$E_{\mathrm{SN}}$ the kinetic energy of the ejecta from each supernova
($\approx 10^{51} \mathrm{erg}$) and $\epsilon_{\mathrm{thermal}}$ the
efficiency with which supernova energy is deposited in the gas, which
is highly uncertain.
With equations \ref{mrh} and \ref{Etherm} $\epsilon^{\mathrm{SN}}_0$
is defined as:
\begin{equation}
\epsilon^{\mathrm{SN}}_0 =  \epsilon_{\mathrm{thermal}}\ \frac{
  \eta_{\mathrm{SN}}\ E_{\mathrm{SN}}}{(200 \mathrm{km/s})^2},
\end{equation}
 We assume a value for the thermal efficiency of
 $\epsilon_{\mathrm{thermal}} \approx 0.16$.

Additionally we assume to have a kinetic SN feedback,
    i.e. reheated gas can be blown out of the halo. Thereby, the
    fraction of reheated gas, which is 
ejected from the halo into the intergalactic medium (IGM), is given by
\begin{equation}
f_{\mathrm{eject}}(V_{\mathrm{vir}}) = \left[1+ \left(
    \frac{V_{\mathrm{vir}}}{V_{\mathrm{eject}}}
  \right)^{\alpha_{\mathrm{eject} 
}} \right]^{-1}
\end{equation}
with $\alpha_{\mathrm{eject}} = 6$ and $V_{\mathrm{eject}}$ a free
parameter ($\approx 100-150\, \mathrm{km/s}$). 
Note that using this definition the total amount of released energy
from SN explosions is not exceeded.

Moreover, the ejected gas can
re-collapse onto the halo at later times and then is available for
cooling. As in \citet{Springel01} and \citet{DeLucia07} the rate of
reinfall of rejected gas is given by
\begin{equation}
\dot{m}_{\mathrm{reinfall}} = \chi_{\mathrm{reinfall}} \left(
  \frac{m_{\mathrm{eject}}}{t_{\mathrm{dyn}}} \right) 
\end{equation}
Here, $\chi_{\mathrm{reinfall}}$ is a free parameter,
$m_{\mathrm{eject}}$ is the mass of the ejected gas outside of the
halo and $t_{\mathrm{dyn}} = r_{\mathrm{vir}}/V_{\mathrm{vir}}$ is the
dynamical time of the halo.  
This treatment is quite similar to that used in simulations which
  explicitly include large-scale winds (e.g. \citealp{Oppenheimer08}),
  however, as noted above the simulations considered here do not
  include such a treatment of winds, and the thermal energy deposition
  that is included is not able to drive gas out of the galaxies. 
}

\item{{\bf{Metal enrichment}}: To track the production of metals, we
  assume that, together with a parcel of new stars $\mathrm{d}m_*$ a
  certain mass of metals $\mathrm{d}M_Z = y \mathrm{d}m_*$ is created
  and instantaneously mixed with the cold gas in the disc. The yield
  is assumed to be constant and is treated as free parameter. Whenever
  new stars are formed, they are assumed to have the metallicity of
  the cold gas at this time step. When metals get ejected from the
  disc due to SN-winds, either the metals are mixed with the hot gas
  or ejected from the halo into the `diffuse' IGM in the same
  proportion as reheated cold gas. Note that only metal enrichment due
  to Supernovae TypeII is tracked. Note that in our simulations
      we consider only primordial metallicity cooling and no metal
      evolution is included.}

\item{{\bf{Black hole growth and AGN feedback}}: Every ``top level''
  halo (halo with no progenitors) in the merger tree is seeded with a
  black hole with mass $\sim 100 M_{\odot}$. Black holes can grow by
  two channels: quasar mode and radio mode. The quasar mode is the
  bright mode of black hole growth observed as optical or X-ray bright
  AGN radiating at a significant fraction of their Eddington limit ($L
  \approx (0.1-1)L_{\mathrm{Edd}}$; \citealp{Vestergaard04,
    Kollmeier06}).  Such bright AGN are believed to be fed by
  optically thick, geometrically thin accretion disks
  (\citealp{Shakura73}). In contrast, AGN activity in the radio-mode
  is much less dramatic. A large fraction of massive galaxies are
  detected at radio wavelengths (\citealp{Best07}) without showing
  characteristic emission lines of classical optical or X-ray bright
  quasars (\citealp{Kauffmann08}).  Their accretion rates are believed
  to be a small fraction of the Eddington rate and they are radiatively
  extremely inefficient. Even if AGN spend most of their time in the
  radio-mode, they gain most of their mass during the short and
  Eddington limited episodes of quasar phases which in the model are
  assumed to be triggered by merger events. The energy released during
  the rapid growth of the black holes can drive powerful galactic
  scale winds that sweep cold gas out of the galaxy.

In contrast to the quasar mode, the radio mode has low-Eddington ratio
accretion rates, is radiatively inefficient and associated with
efficient production of radio jets that can heat gas in a
quasi-hydrostatic hot halo.  Assuming Bondi-Hoyle accretion combined
with an isothermal cooling flow solution (\citealp{Nulsen00}) we
calculate the accretion rate in the radio mode
\begin{equation}
\dot{m}_{\mathrm{radio}} = \kappa_{\mathrm{radio}} \left[
  \frac{kT}{\Lambda(T, Z_h)} \right] \left( \frac{M_{\bullet}}{10^8
  M_{\odot}} \right). 
\end{equation}
Note that the central black hole accretes at this rate whenever hot
halo gas is present (`hot mode' accretion,
$r_{\mathrm{cool}}<r_{\mathrm{vir}}$). The energy that effectively
couples to and heats the hot gas is given by $L_{\mathrm{heat}} =
\kappa_{\mathrm{heat}} \eta_{\mathrm{rad}} \dot{m}_{\mathrm{radio}}
c^2$.  Assuming that all the hot gas is at the virial temperature of
the halo, the heating rate is given by
\begin{equation}
\dot{m}_{\mathrm{heat}} = \frac{L_{\mathrm{heat}}}{3/4\ V^2_{\mathrm{vir}}}
\end{equation}
The net cooling rate is then the usual cooling rate minus the heating
rate from the radio-mode. Note that in our simulations, black hole
    growth and AGN feedback is not implemented.}

\end{enumerate}

Note that in merger trees from N-body simulations it may happen that
the total mass of two merging halos at the beginning of a merger event
is larger than the mass of the merged object afterwards, as during the
merger particles can become unbound through tidal forces. Therefore,
in the SAM we impose an upper limit on the hot halo mass of
\begin{equation}
M_{\mathrm{hot}} = M_{\mathrm{bar}} - M_{\mathrm{star,tot}} -
M_{\mathrm{cold}} -  M_{\mathrm{eject}} 
\end{equation}
Here, $M_{\mathrm{eject}}$ is the mass ejected by winds,
$M_{\mathrm{star,tot}}$ and $M_{\mathrm{cold}}$ are the total star and
cold gas masses within the merged halo and $M_{\mathrm{bar}}$ is the
expected baryonic fraction of the halo. In this way, we prevent the
sum of all baryonic components in the halo from exceeding the
universal baryon fraction.

\section{Redshift evolution of galaxy properties}\label{galprop} 

In this section, we compare the cosmic evolution of the baryonic
components of the galaxies and halos from the direct cosmological
simulations to the results from the SAMs using the dark matter merger
trees constructed from the re-simulations. Here we only consider the
evolution of the central galaxy in the main branch of the merger tree
(largest progenitor halo). We divided the $48$ halos into three bins
according to their halo mass at $z=0$ (every bin contains $16$ halos)
with $4.5 \times 10^{12} M_{\odot} < M_{\mathrm{halo}} < 4 \times
10^{13} M_{\odot}$ (high-mass), $1.2 \times 10^{12} M_{\odot} <
M_{\mathrm{halo}} < 4.5 \times 10^{12} M_{\odot}$ (intermediate-mass),
and $2.4 \times 10^{11} M_{\odot} < M_{\mathrm{halo}} < 1.2 \times
10^{12} M_{\odot}$ (low-mass). All comparisons in this Section make
use of these bins. 

Note also that a resolution study of the evolution of the baryonic
component in SAMs based on $2\ \times$ and $4\ \times$ higher
resolution simulations for a high- and a low mass halo can be found in
the Appendix. In both cases the results based on the simulations with 
different resolution are consistent.

\subsection{Baryon fraction}\label{barfrac} 

\begin{figure}
\begin{center}
  \epsfig{file= 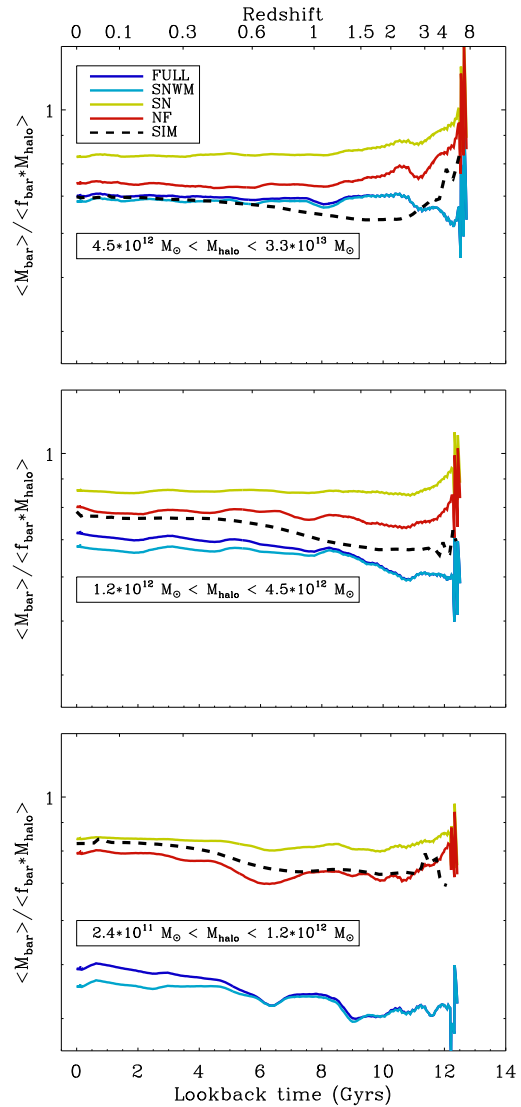, width=0.45\textwidth}
  \caption{Total baryonic mass $M_{\mathrm{bar}} =
    M_{\mathrm{gas}}+M_*$, as a fraction of the cosmic baryon fraction
    times the halo mass $f_{\mathrm{bar}}\times M_{\mathrm{halo}}$, as
    a function of lookback time for different semi-analytic models (red:
    no feedback - NF; green: thermal Supernova feedback - SN; blue:
    thermal Supernova feedback, SN-driven winds and metal cooling -
    SNWM; purple: `full' model including feedback from black holes -
    FULL) and for the SPH-re-simulation (black lines). Upper panel:
    Average values for the high mass bin with $4.5 \times 10^{12}
    M_{\odot}< M_{\mathrm{halo}} < 3.3 \times 10^{13}
    M_{\odot}$. Middle panel: Average values for the intermediate mass
    bin with $1.2 \times 10^{12} M_{\odot}< M_{\mathrm{halo}} < 4.5
    \times 10^{12} M_{\odot}$. Lower panel: average values for halo
    masses between $2.4 \times 10^{11} M_{\odot}< M_{\mathrm{halo}} <
    1.2 \times 10^{12} M_{\odot}$. }  {\label{mah_sch}}
\end{center}
\end{figure}
For a first comparison we compute the total baryonic mass
$M_{\mathrm{bar}} = (M_{\mathrm{star}} + M_{\mathrm{cold}} +
M_{\mathrm{hot}})$ within the virial radius of the main halo at every
redshift for the simulations and the SAMs, respectively. In the
re-simulations as well as in the SAMs, we consider the hot gas mass
within the whole halo (i.e. within the virial radius), but the stars
and the cold gas only of the central galaxy. We neglect contributions
from substructures, diffuse stars and cold gas and satellite stars and
cold gas.  The baryonic mass is compared to the mass of available
baryons within each halo, defined as $f_{\mathrm{bar}} \times
M_{\mathrm{halo}}$, where $f_{\mathrm{bar}} = 0.169$ is the cosmic
baryon fraction.

In Fig. \ref{mah_sch} we show the average ratio of
$M_{\mathrm{bar}}/(f_{\mathrm{bar}} \times M_{\mathrm{halo}})$, as a
function of redshift for the three mass bins. The simulations are
compared to the four SAM variants: NF, SN, SNWM, and FULL (for details
see section \ref{sam}). We expect the simulations to be most directly
comparable to the NF or SN SAMs, as these SAMs include the same
complement of physical processes as the simulations. Considering first
the NF model, we see that at low redshift, the SAM overestimates the
baryon fraction in high mass halos, nearly agrees in intermediate mass
halos, and slightly underestimates it in low mass halos. At high
redshift, the NF SAM overestimates the baryon fraction at high
redshift in high and intermediate mass halos, by a somewhat larger
factor in the former. Turning next to the SN SAM, we see that the SAM
predicts baryon fractions that are everywhere higher than the
simulation results, though much more so for the high and intermediate
mass halos.
This is because the ``thermal'' SN feedback removes baryons from
satellite galaxies, which are not counted in this census, and
deposits them in the hot gas component which is included here.
In the SNWM and FULL model, we see the impact of the SN-driven
winds, which remove baryons from low-mass halos. The
additional metal cooling in the SNWM and the FULL model does not
have any impact on the total baryon fraction (only on the individual
components, see next sections.)
AGN feedback (in the FULL model) mainly prevents hot gas from cooling,
so does not affect the total baryon fraction significantly, but will
be important for the fraction of stars and hot gas, which will be
discussed later. 

\subsection{Cold gas and stars}\label{stellarevol}

\begin{figure}
\begin{center}
  \epsfig{file= 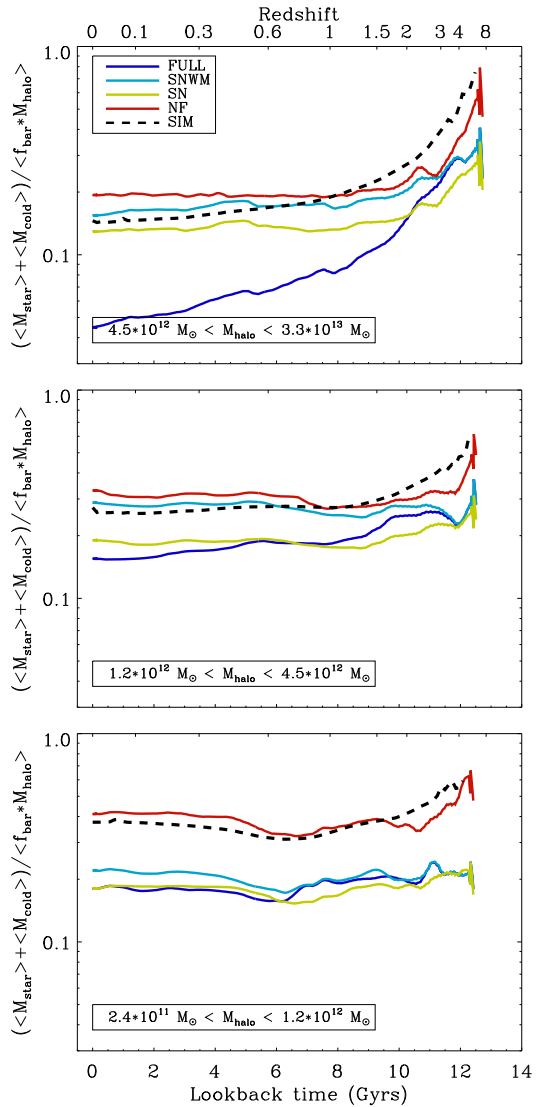,
    width=0.45\textwidth}  
  \caption{The evolution of the mass of the condensed baryons
    $M_*+M_{\mathrm{cold}}$ as a fraction of the total baryon mass as
    a function of redshift for the three mass bins. The condensed
    baryon fraction in halos of all masses is in reasonable agreement
    with the NF model at all redshifts. We can see that SN FB most
    strongly affects the low-mass bin, while AGN FB affects the high
    mass bin.}  {\label{mah_sc}}
\end{center}
\end{figure}

In Fig.~\ref{mah_sc} we show the evolution of the mass of condensed
baryons (stars and cold gas, $M_*+M_{\mathrm{cold}}$) as a fraction of
the total baryon mass as a function of redshift for the three mass
bins. There is fairly good agreement between the simulations and the
NF model over the whole redshift range, although the SAM is a
little low at high redshift and a bit high at low redshift,
particularly in the high and intermediate mass bin. In the SN model,
the condensed baryon fraction is lowered by an almost fixed factor
relative to the NF model, and is significantly lower than the
simulation results. This suggests that the ``thermal feedback''
implemented in the simulation is less effective than that included in
the SAM. 

It may seem curious that the SNWM model results are \emph{higher}
    than the SN model, in fact close to the NF model in the high and
    intermediate mass bin. This is because the SNWM model includes
    metal cooling, leading to more efficient star formation, while the
    SN model does not. The enhanced cooling rates partly compensate
    for the removal of cold gas via the SN-driven winds. In particular
    for the high- and intermediate mass bin the effect of the SN winds
    is almost the opposite of that of metal cooling, resulting in
    condensed baryon fractions close to the NF model. Only in the
    low-mass bin, the effect of SN winds is not completely compensated
    by metal cooling (so condensed baryon fractions are below the NF
    model).

Finally, we see that in the FULL model, the AGN FB begins to
quench star formation in the massive halos after about $z\sim 2$,
while it has little effect on the lower mass bins.

\begin{figure}
\begin{center}
  \epsfig{file= 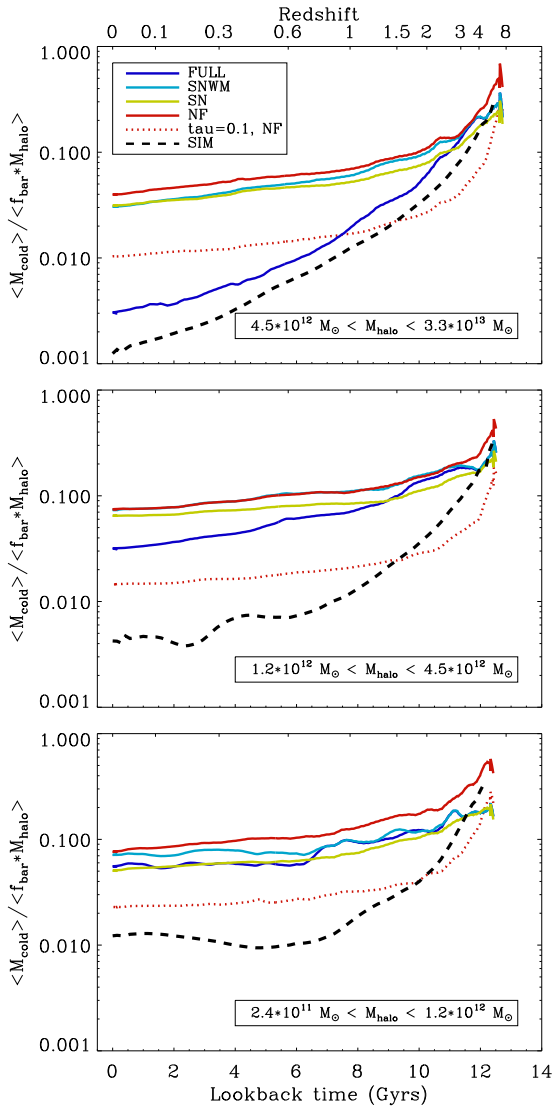, width=0.45\textwidth}
  \caption{Evolution of mean cold gas fraction
    $M_{\mathrm{cold}}/(f_{\mathrm{bar}}\times M_{\mathrm{halo}})$ of
    central galaxies. Mass bins and colors are the same as in
    Fig. \ref{mah_sch}. In all simulations the cold gas is 
  depleted more efficiently than in the SAMs due to the large star
  formation efficiency at high redshifts. The red dotted line shows
  the cold gas fraction assuming ten times higher efficiency for star
  formation in the NF model (see Eq. \ref{norm}).}  
 {\label{mah_cold}}
\end{center}
\end{figure}
In Fig.~\ref{mah_cold} we plot the evolution of the mean cold gas
fraction of the central galaxy (in Fig.~\ref{mah_sc} we plotted cold
gas plus stars). The efficiency of the conversion of cold gas to stars
is clearly very different between the simulations and SAMs. In both
cases (simulations and SAMs), the final cold gas fraction is
increasing with decreasing halo mass. For the NF model, the SN and the
SNWM model the cold gas fraction varies only slightly over time and is
significantly higher (about an order of magnitude since $z=2$) than
for the simulations. This shows that the inclusion of SN FB has little
impact on the gas fractions of galaxies. Only the FULL model shows a
much stronger decrease of the gas fraction with cosmic time for
massive galaxies due to the radio mode feedback. The initial cold gas
fraction, at high redshifts $4<z<8$, is almost the same for the
simulations and SAMs. With evolving cosmic time the cold gas content
decreases more rapidly in the simulations due to the more efficient
conversion into stars. The cold gas in the simulations is already
converted into stars at high redshift and there is almost no more cold
gas left to turn into stars at lower redshifts. This is similar to the
results found in the comparison of \citet{Cattaneo07}.
\begin{figure}
\begin{center}
  \epsfig{file= 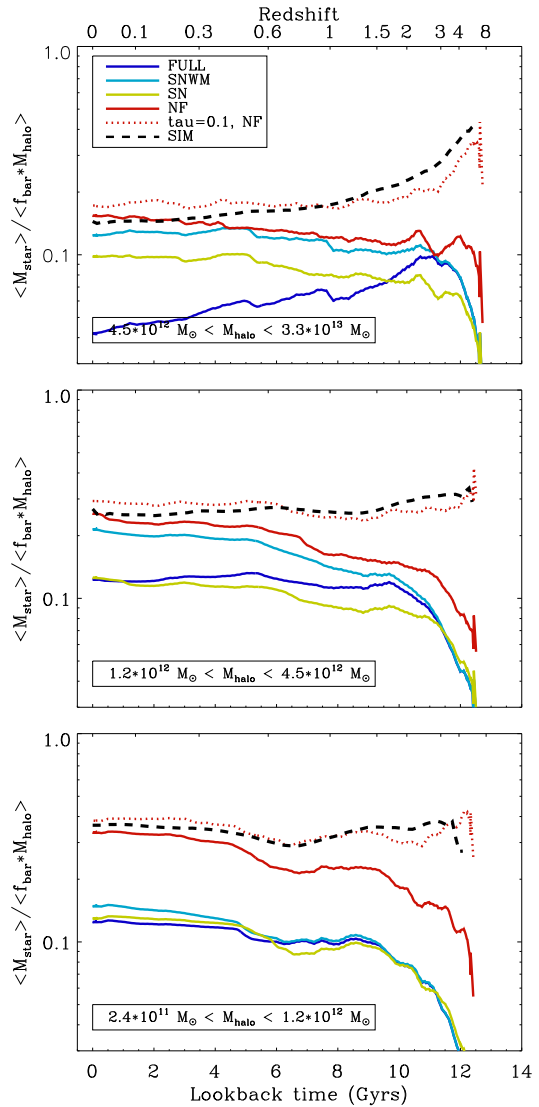, width=0.45\textwidth}
  \caption{Comparison of the stellar baryon fraction of the central
    galaxy between simulations and SAMs. The mass bins and colors are
    the same as in Fig. \ref{mah_sch}. For all mass bins the
    simulations agree best with the NF SAM, but form significantly
    more stars at high redshifts $z > 1$. We can reproduce the
    simulation results fairly well with the NF SAM if we increase the
    star formation efficiency parameter by a factor of ten (red dotted
    lines). At high masses the AGN feedback (FULL) and at low
    masses the SN feedback (SN and SNWM) reduce the stellar baryon
    fractions in the SAMs.}  {\label{mah_stars}}
\end{center}
\end{figure}

In Fig.~\ref{mah_stars} we show the corresponding fraction of
available baryons that are converted into stars in the central galaxy
$M_{\mathrm{*}}/(f_{\mathrm{bar}} \times M_{\mathrm{halo}})$,
sometimes termed baryon conversion efficiency \citep{Guo09,Moster10}.
In general, all simulations predict a decreasing (high-mass) or almost
constant conversion efficiency with redshift (low-mass), whereas most
SAMs predict increasing conversion efficiencies with the exception of
high-mass galaxies in the FULL model with AGN feedback.

At low redshift $z<0.6$ the conversion efficiencies agree well between
the simulations and the NF model, with higher values for lower mass
galaxies. However, at high redshifts $z>1$ the conversion efficiencies
are significantly higher for the simulations. This is in contrast to
the results of \citet{Cattaneo07}, where the stellar masses agree at
high redshift, but the SAM masses are larger than in simulations at
low redshift. The difference in the behaviour of the SAMs and the
simulations can be explained in terms of star formation efficiency. We
changed the normalization of the SK-relation in the SAM by introducing
a factor $\tau_*$ in Eq. \ref{SK}:
\begin{equation}\label{norm}
\Sigma_{\mathrm{SFR}} = \frac{A_{\mathrm{KS}}}{\tau_*} \Sigma_{\mathrm{gas}}^{N_K},
\end{equation}
with $\tau_* \approx 0.1$. The results of the NF model with this
elevated star formation efficiency for the stellar and cold gas mass
evolution is shown in Figs.~\ref{mah_cold} and \ref{mah_stars}. At
high redshift, gas is more efficiently depleted and converted into
stars, resulting in a better agreement between this `high SFE' model
and the simulations. However, for the high-mass and intermediate-mass
bin, the `high SFE' model overpredicts the stellar fraction at low
redshifts, suggesting that the \citet{Cattaneo07} SAMs may also have
had a higher SFE, and this could explain the discrepancy between their
results and our initial results.
In all three mass bins, the NF model produces the most massive stellar
components. Again, the more efficient cooling due to metals in the
SNWM and the FULL model is cancelled by the effect of winds and, for
massive galaxies, also by AGN feedback, resulting in lower stellar
masses than in the NF model. The separate behaviour of metal cooling
and SN winds is the same as for the total condensed baryon fraction.

\subsubsection{Star Formation Rates}

\begin{figure}
\begin{center}
  \epsfig{file= 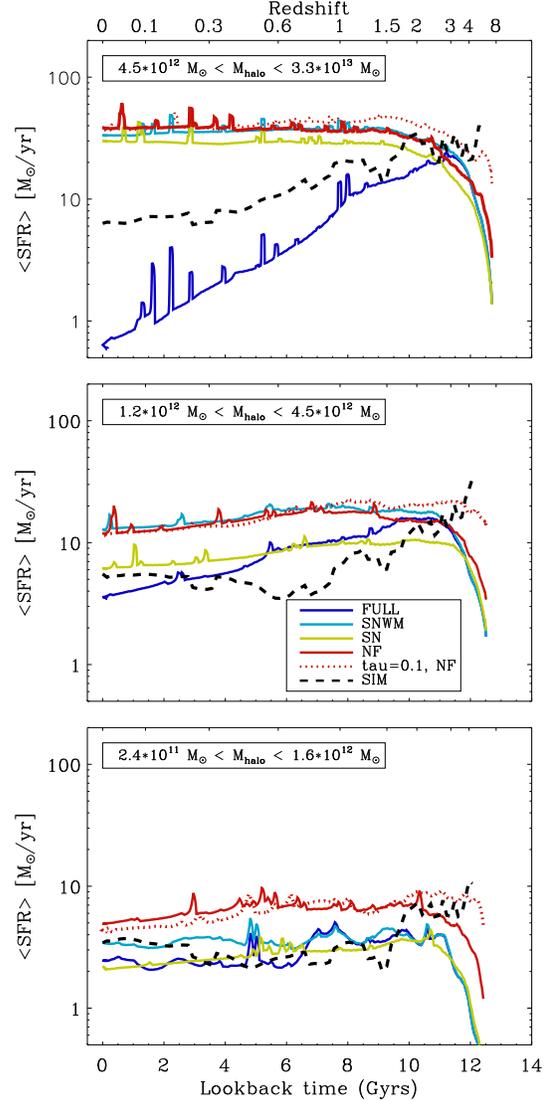, width=0.45\textwidth}
  \caption{Evolution of the star formation rates in simulations and
    SAMs for three mass bins. The mass bins and colors are the same as
    in Fig. \ref{mah_sch}. At high redshifts, SFRs are higher in the
    simulations than in the SAMs, leading to more rapid depletion of
    the cold gas and very low SFRs at low redshifts.}
          {\label{mah_sfr}}
\end{center}
\end{figure}

To confirm the previous findings we compare the star formation rates
in Fig. \ref{mah_sfr}. At very high redshifts $z>4$, the SFRs in the
simulations are much higher than in the SAMs. Only by assuming more
efficient star formation in the NF SAM ($\tau_*=0.1$) do we obtain a
reasonable match to the simulations.  However, at $z<1.5$, the high
SFE model results in similar SFRs as the original NF model, as the
larger SF efficiencies at high redshift lead to a more rapid depletion
of the cold gas. In the simulation, the cold gas is rapidly turned
into stars, resulting in lower SFRs at low redshifts compared to SAMs
because of gas depletion. Only the FULL model shows a strongly
decreasing SFR with decreasing redshift due to radio mode feedback,
which becomes especially important for low redshifts and large halo
masses. This result is consistent with the study of \citet{Saro10},
who compared their stripped-down versions of SAMs (with no feedback)
to simulations, and found find higher SFRs in the simulations for all
galaxies within a cluster (central and satellites) at high redshifts
and lower SFRs at low redshifts. In addition, \citet{Stringer10} find
a similar discrepancy for the specific star formation rates at high
redshifts (larger in simulations than in their SAM) and good agreement
for low redshifts.

\begin{figure}
\begin{center}
  \epsfig{file= 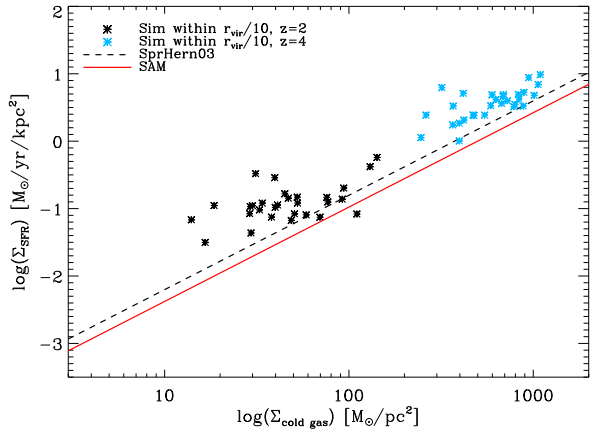, width=0.45\textwidth}
  \caption{Star formation rate surface densities versus cold gas
    surface densities for the simulated galaxies within $1/10
    r_{\mathrm{vir}}$. Black and blue stars correspond to different
    re-simulations at $z=2$ or $z=4$, respectively. The red solid line
    illustrates the Kennicutt relation implemented in the SAM with a
    normalization consistent with a Chabrier IMF, while the black
    dashed line the one used to normalize the simulations
    (\citealp{Springel03}), which assumes a Salpeter IMF. }
          {\label{SurfDens}}
\end{center}
\end{figure}

To better understand this discrepancy between simulations and SAMs we
take a closer look at the respective implementations of star
formation. According to \citet{Springel03}, stars in the simulations
are formed locally out of cold gas with the star formation rate
density proportional to the local three-dimensional density of gas to
the power of 1.4, $\rho_{\mathrm{SF}} \propto \rho^{1.4}/t_*$. The
star formation timescale $t_*$ was set to reproduce the observed local
Schmidt-Kennicutt relation (SK) for a simulation of a smooth, isolated
disk-dominated galaxy set up to resemble the Milky Way. In the SAMs
the cold gas is assumed to settle into smooth exponential disks and
stars form according to the SK-relation, implemented in terms of
surface densities.

In Fig.~\ref{SurfDens}, we plot the SFR surface density versus the
surface density of the cold gas for the simulated galaxies at $z=2$
and $z=4$ within $1/10\ r_{\mathrm{vir}}$ for all re-simulations. The
black dashed line is the SK-relation assuming a Salpeter IMF
\citep{1955ApJ...121..161S}, as given in the original Kennicutt papers
and as implemented in the simulations following \citet{Springel03}. We
would naively expect the simulations to follow this line. The red
solid line shows the SK-relation for a Chabrier IMF
\citep{2003PASP..115..763C}, as assumed in the SAMs. At a given gas
surface density, the SFR surface densities of the simulations lie
mostly above the expected SK-relation. The change of normalization
associated with converting from Salpeter to Chabrier cannot account
for the increased star formation efficiency in the simulations. In
general, star formation in the cosmological simulations is about a
factor of five 
more efficient than for simulations of smooth isolated
disks using the identical model (see \citealp{Springel03}). This
discrepancy is a consequence of the \textit{clumpy} structure of cold
gas in the cosmological simulations. In the clumps the gas can reach
higher local densities than in the idealized smooth disks that have
been used by \citet{Springel03} to calibrate the star formation
timescale by matching the SK-relation. As the implemented SK-relation
is not linear, the structure of the cold gas distribution plays an
important role for the overall star formation efficiency within the
galaxies (see \citet{2010ApJ...720L.149T} for a discussion on galaxy
mergers). In other words, for any star formation model with a
non-linear dependence on the local gas density (exponent larger than
unity), a more clumpy gas distribution will effectively increase the
star formation efficiency. These combined effects explain the much
higher SF efficiencies at high redshift in the simulations relative to
the SAMs.

\begin{figure}
\begin{center}
  \epsfig{file= 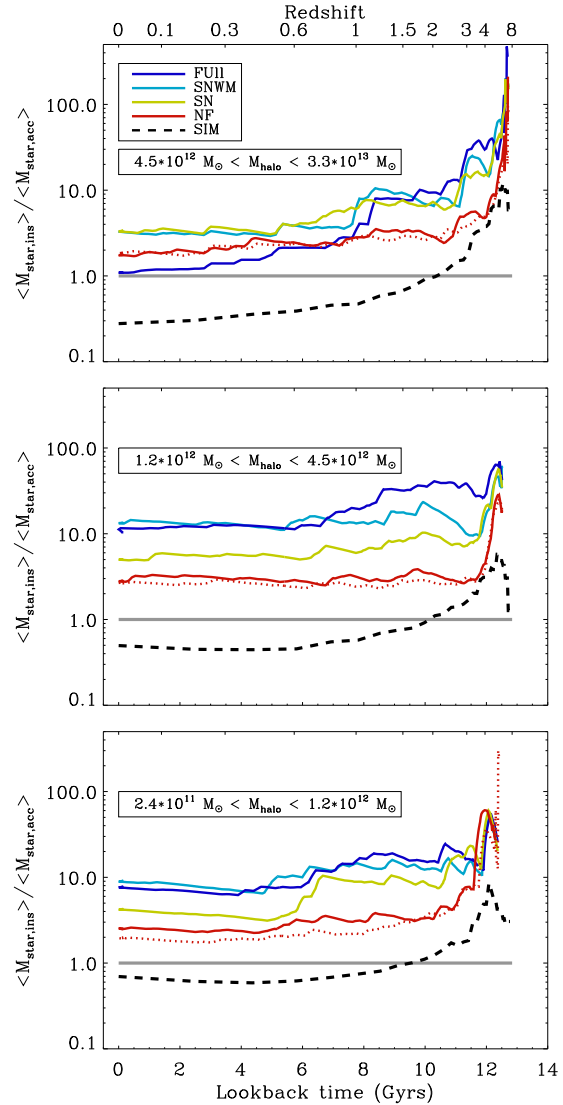, width=0.45\textwidth}
  \caption{Fraction of in situ to accreted stellar mass versus
    redshift for different halo masses.  The mass bins and colors are
    the same as in Fig. \ref{mah_sch}. 
In situ star formation dominates over accretion in all the SAM
variants at all redshifts, in contrast to the simulations for which
accretion dominates at late times, especially in massive halos.}
          {\label{mah_frac}}
\end{center}
\end{figure}

\subsubsection{Modes of Stellar Mass Growth}

\begin{figure*}
\begin{center}
  \epsfig{file= 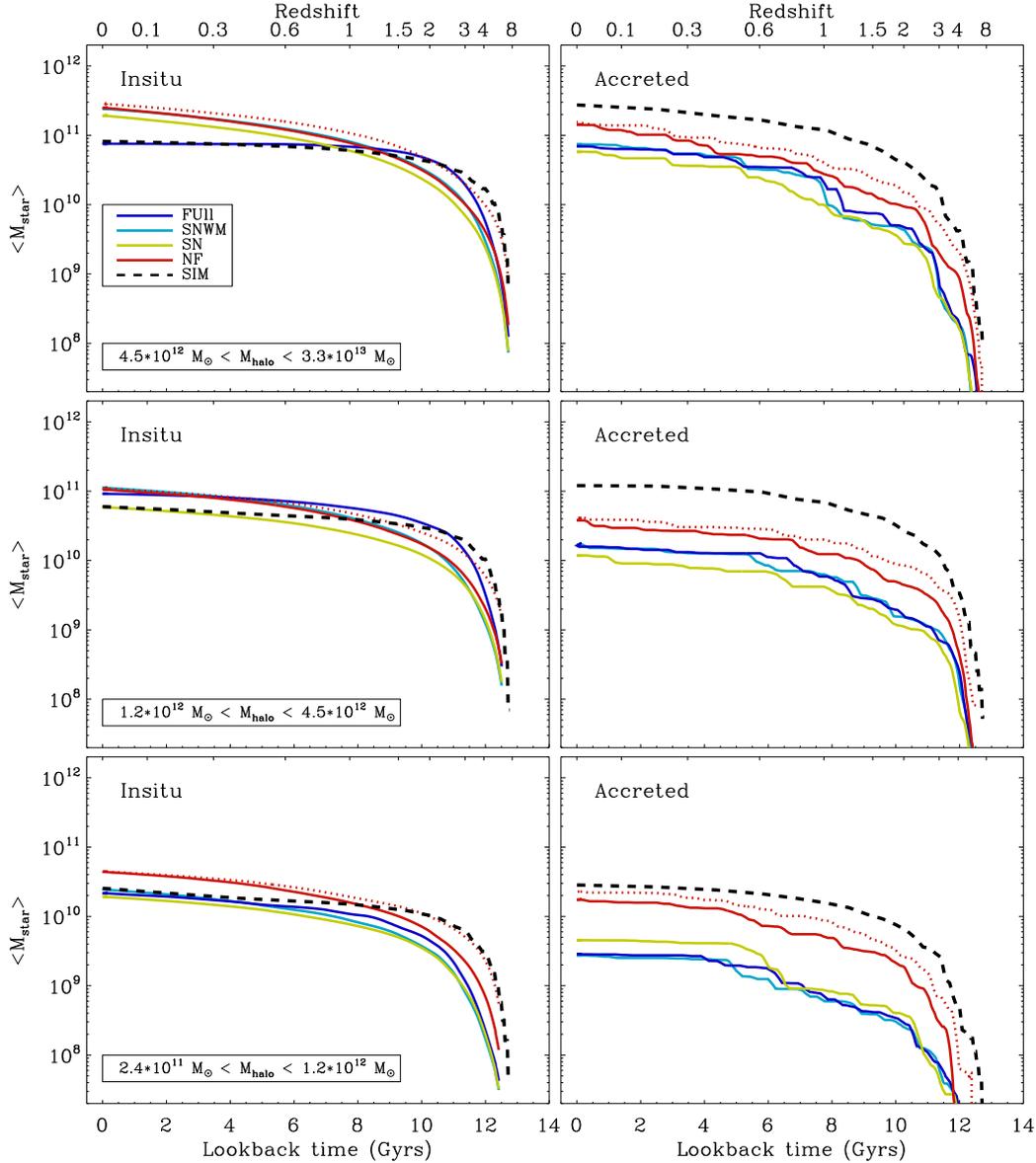, width=0.8\textwidth}
  \caption{Comparison of the mean in situ (left column) and accreted
    (right column) stellar masses in different mass bins. The mass
    bins and colors are the same as in Fig. \ref{mah_sch}. The in situ
    stellar masses in SAMs agree with the ones in simulations
    reasonably well, whereas the accreted stellar mass is smaller in
    the SAMs than in the simulations.}  {\label{mah_accins}}
\end{center}
\end{figure*}

In the hierarchical picture, galaxies can grow their stellar masses in
two ways: 1) by converting cold gas into stars in situ 2) by accreting
already formed stars via mergers. We refer to these two modes as
``in situ'' and ``accreted''.  Our simulations exhibit two phases of
growth, with a rapid early phase at $z>2$ during which stars are
formed in situ from infalling cold gas, followed by an extended phase
at $z<3$ during which the growth is primarily due to accretion of
stars formed in external galaxies (\citealp{Oser10}). 
We now investigate whether the SAMs show the same behavior. In
Fig. \ref{mah_frac}, we show the fraction of cumulative in situ over
accreted stellar mass as a function of redshift for the three
different mass bins. For the SAMs the qualitative trend of a
decreasing fraction of in situ growth is reproduced for the high mass
bin. However, the fraction of in situ formed stars dominates over
accreted stars for all models, all masses, and at all
redshifts. This is in contrast with the simulations, where accretion
dominates over in situ formation for massive systems at low redshifts
as discussed in \citet{Oser10}.

We note several interesting trends in the in situ to accreted fraction
as we vary the physics in the SAMs. Adding thermal SN FB increases the
in situ fraction at all redshifts and in all mass bins. This is
presumably because it suppresses star formation in low-mass satellites
which are the source of accreted stars. Adding the SN-driven winds and
metal cooling further increases the in situ fraction, again at all
redshifts below $z \sim 4$. Switching on AGN FB increases the in situ
fraction at high redshift and decreases it at low redshift in the high
mass bin (and to a lesser extent in the intermediate mass bin). This
is because the radio mode feedback shuts off cooling at late times in
massive halos, removing the supply of new gas needed to fuel ongoing
in situ star formation. Interestingly, increasing the star formation
efficiency in the NF model has almost no effect on the in situ to
accreted fraction. This is presumably because the SF efficiency is
increased in the central and (accreted) satellite galaxies
alike. However, if the SFE were higher in high redshift galaxies than
at low redshift, this would presumably increase the accreted fraction
in present day galaxies. This may be part of the reason for the higher
accreted fractions in the simulations.

In Fig.~\ref{mah_accins}, we show the evolution of the cumulative mass
of insitu and accreted stars separately for the simulations and the
various SAM variants.  Here we can see that the NF SAM actually
reproduces the growth of \emph{in situ} stellar mass fairly well,
though overproducing the in situ mass at low redshift somewhat,
especially in the highest mass bin. We speculate that gravitational
heating in the simulations prevents some of the late cooling in the
highest mass bin and leads to lower in situ stellar mass than the NF
SAM
\citep{2007ApJ...658..710N,2009ApJ...699L.178N,2009ApJ...697L..38J, Feldmann10}.
The radio mode AGN FB in the FULL model leads to a similar suppression
of this in situ mass growth in the massive halos. The NF model with
increased SFE gives an even better match to the simulations at high
redshift. The discrepancy arises from the much lower accreted masses
in the SAM. Here again, the SF model with high SFE comes the closest
to matching the simulation results, but it still falls short by a
considerable amount.

Part of the reason for the lower predicted accreted masses in the SAMs
is that the SAMs used here only allow cooling onto the central galaxy
in the halo, effectively assuming that the hot gas reservoir of a
satellite galaxy is stripped as soon as it enters the virial radius of
the host. This is known to result in satellites that are too red and
have star formation rates that are too low compared with observations
\citep{kimm:09}. It will also truncate their star formation, resulting
in a smaller amount of stellar mass that will eventually be accreted
when they merge (\citealp{Khochfar08}).

\subsection{Hot halo gas}\label{hotevol}

\begin{figure}
\begin{center}
  \epsfig{file= 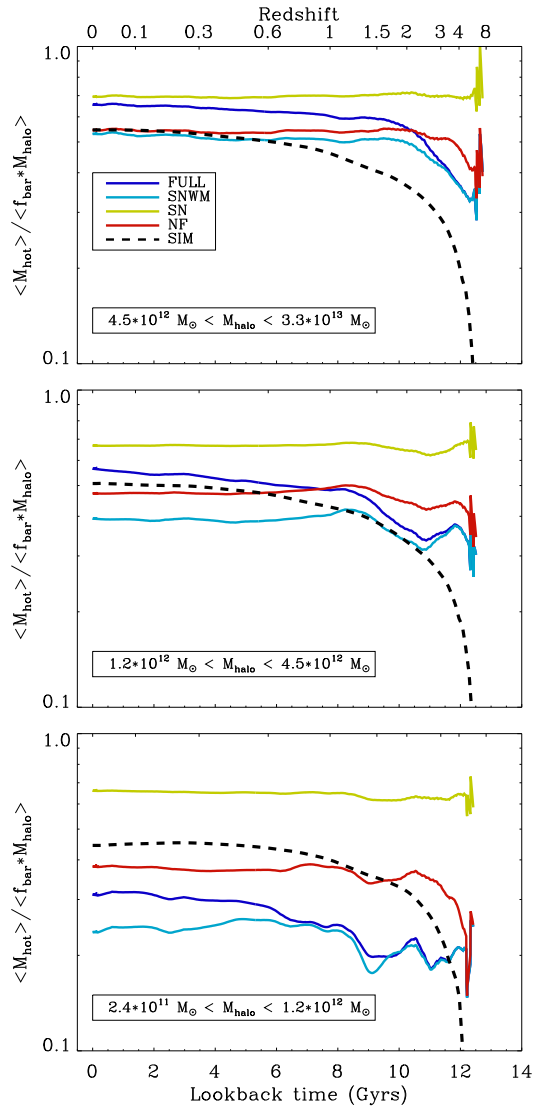, width=0.45\textwidth}
  \caption{Evolution of the mass fraction of hot gas in simulations
    and SAMs normalized to the available mass in baryons. The mass
    bins and colors are the same as in Fig. \ref{mah_sch}. The SAMs
    tend to overpredict the mass of gas at high redshift relative to
    the simulations. }  {\label{mah_hot}}
\end{center}
\end{figure}

\begin{figure*}
\begin{center}
  \epsfig{file= 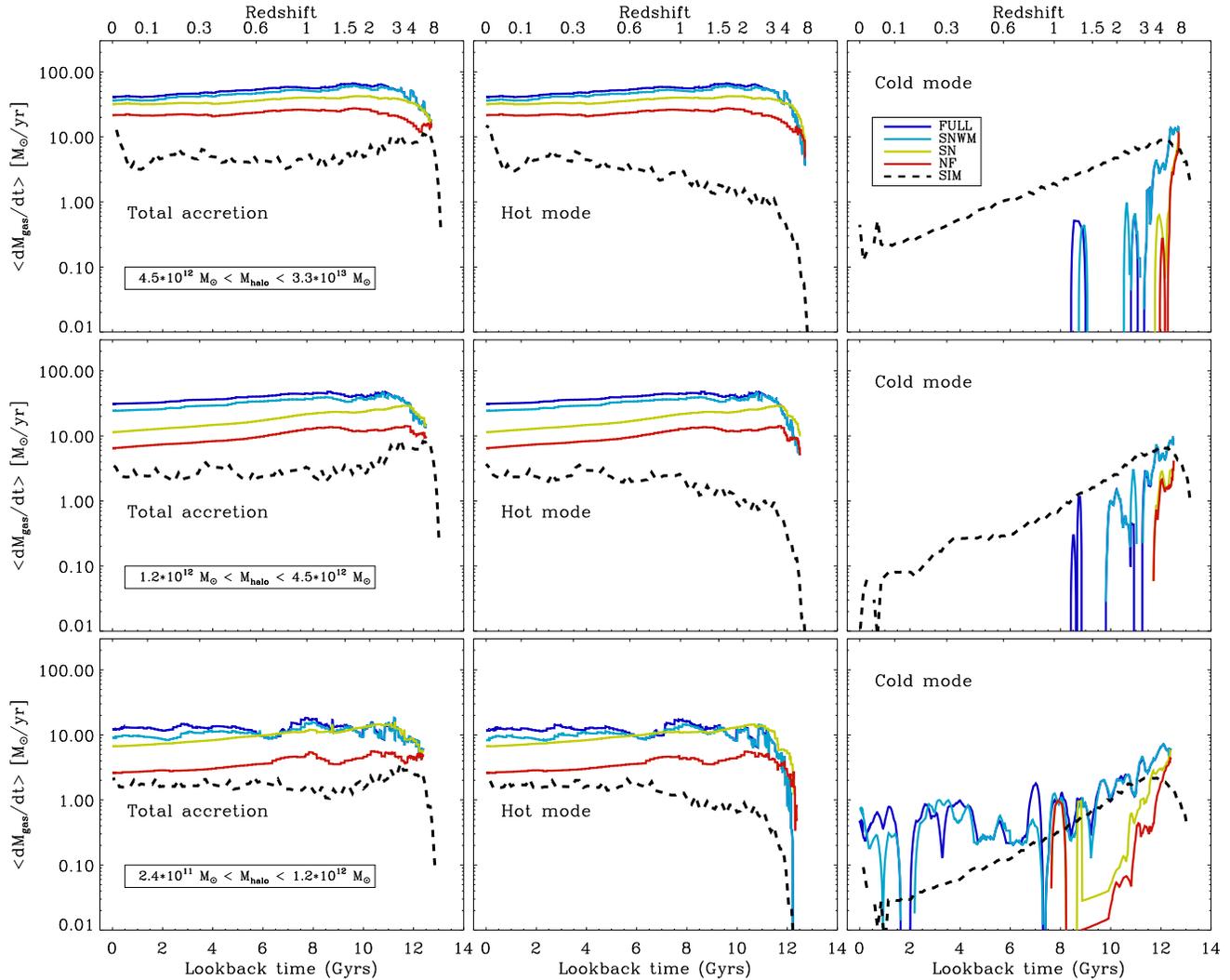, width=1.0\textwidth}
  \caption{Comparison of the mean accretion rates of all gas (left
    column), the rate of hot mode accretion (intermediate column) and
    of cold mode accretion (right column) in simulations and
    SAMs onto the central galaxies. The mass bins and colors are the
    same as in Fig. \ref{mah_sch}. Note that in the FULL and SNWM we
    have not substracted heating rates or rates from blown-out gas
    from the accretion rates. Compared to simulations, hot mode
    accretion is overestimated in all SAMs, whereas cold mode
    accretion is in general underestimated. }  {\label{mah_accr}}
\end{center}
\end{figure*}

The evolution of the mean hot gas fraction is shown in the three
panels of Fig.~\ref{mah_hot}. In all but the SN SAMs the hot gas
fraction increases with increasing halo mass, which is qualitatively
similar to the simulations. In the simulations this effect is caused
by shock heating of infalling baryonic material which becomes more
efficient for massive halos
(e.g. \citealp{1977ApJ...211..638S,1977ApJ...215..483B,1978MNRAS.183..341W,
  2003MNRAS.345..349B,2007MNRAS.380..339B,Keres05,Khochfar08,Keres09,
  2009ApJ...697L..38J}). This trend is also seen by the SAMs,
except for the SN model, where the supernova energy input heats most
of the available gas to the virial temperature of the halos, keeping
the hot gas fraction constant independent of halo mass. For the SNWM
and FULL models, the supernova winds drive some of the hot gas out of
the low mass halos. In contrast, the additional metal cooling shows a
negligible effect on the evolution of the hot halo gas.
The additional effect of Radio mode heating (FULL
model), which prevents late cooling in massive halos and therefore
leads to larger amounts of hot gas, is apparent for the intermediate
and high mass galaxies.  The NF model agrees fairly well with the
simulations in all mass bins at $z\lesssim 1.5$ but substantially
overpredicts the amount of hot gas at high redshift. 

To understand the differences in the hot gas content at high redshift
we investigate the gas accretion modes onto the central galaxies. For the
simulations we distinguish between hot and cold accretion by
considering the highest temperature a gas particle had before it was
accreted onto the galaxy, i.e. 1/10th of the virial radius. We use the
same definition to distinguish between between hot and cold gas as
given by equation \ref{coldhot} (similar to \citealp{Keres05}). In the
SAMs, we distinguish between hot and cold mode accretion (slow and
rapid cooling) depending on whether the ratio of the cooling radius to
the virial radius $r_{\mathrm{cool}}/r_{\mathrm{vir}}$ is larger (cold
mode) or smaller (hot mode) than unity \citep{White91}. The
distinction between hot and cold mode accretion approximately
specifies whether the gas was heated to the virial temperature of the
host halo before it was accreted onto the galaxy (hot mode) or was
directly accreted without being heated (cold mode). The cold mode
accretion is meant to represent the `cold flows' recently discussed in
the literature (\citealp{Keres05,Keres09,Oser10,Dekel09}). Note that
for the SNWM and FULL SAMs, we have not substracted the heating rates
or rates of blown-out gas due to feedback processes from the accretion
rates. 

In Fig. \ref{mah_accr} we show the comparison of the total (left
panels), hot (middle panels) and cold mode (right panels) gas
accretion rates onto the central galaxies as a function of redshift. For all
SAMs, the total gas accretion rates onto the galaxies are
significantly higher than for the simulations. This is caused solely
by higher hot mode accretion rates from the generally larger hot gas
reservoir in particular at high redshift (middel panel in
Fig.~\ref{mah_accr} and see Fig. \ref{mah_hot}). The cold mode
accretion rates are much lower in the SAMs than in the simulations,
and in the SAMs without metal cooling (which are more relevant to
compare with these simulations), the cold mode is truncated at
$z\lesssim 4$ for massive halos, $z\lesssim 1$--1.5 for low mass halos,
while it declines smoothly until $z\sim0$ in the simulations. However,
the predicted rates of cold mode accretion are much higher in the SAMs
that include metal cooling.

In contrast to our results, \citet{Cattaneo07} find a reasonably good
match for the evolution of the hot gas content as well as for the hot
and cold mode accretion rates in their SAM version without any
feedback. However, they include metal cooling in their SAM, but not in
their simulations.  \citet{Benson10} compare cold and hot mode
accretion rates from SAMs to simulations, varying the supernova
feedback and conditions for the rapid cooling regime according to
\citet{Birnboim03}. They concluded that cold-mode physics is already
adequately accounted for in SAMs --- but they also used simulations
with only primordial H \& He cooling, but included metal cooling in
their SAMs. \citet{Lu10} assume only H \& He cooling in their SAMs
\textit{as well as} in the simulations (like we do) and find
qualitatively similar results to ours: a discrepancy for the hot halo
gas fraction at high redshift associated with larger hot mode and
smaller cold mode accretion rates in the SAMs than in the simulations.

\begin{figure*}
\begin{center}
  \epsfig{file= 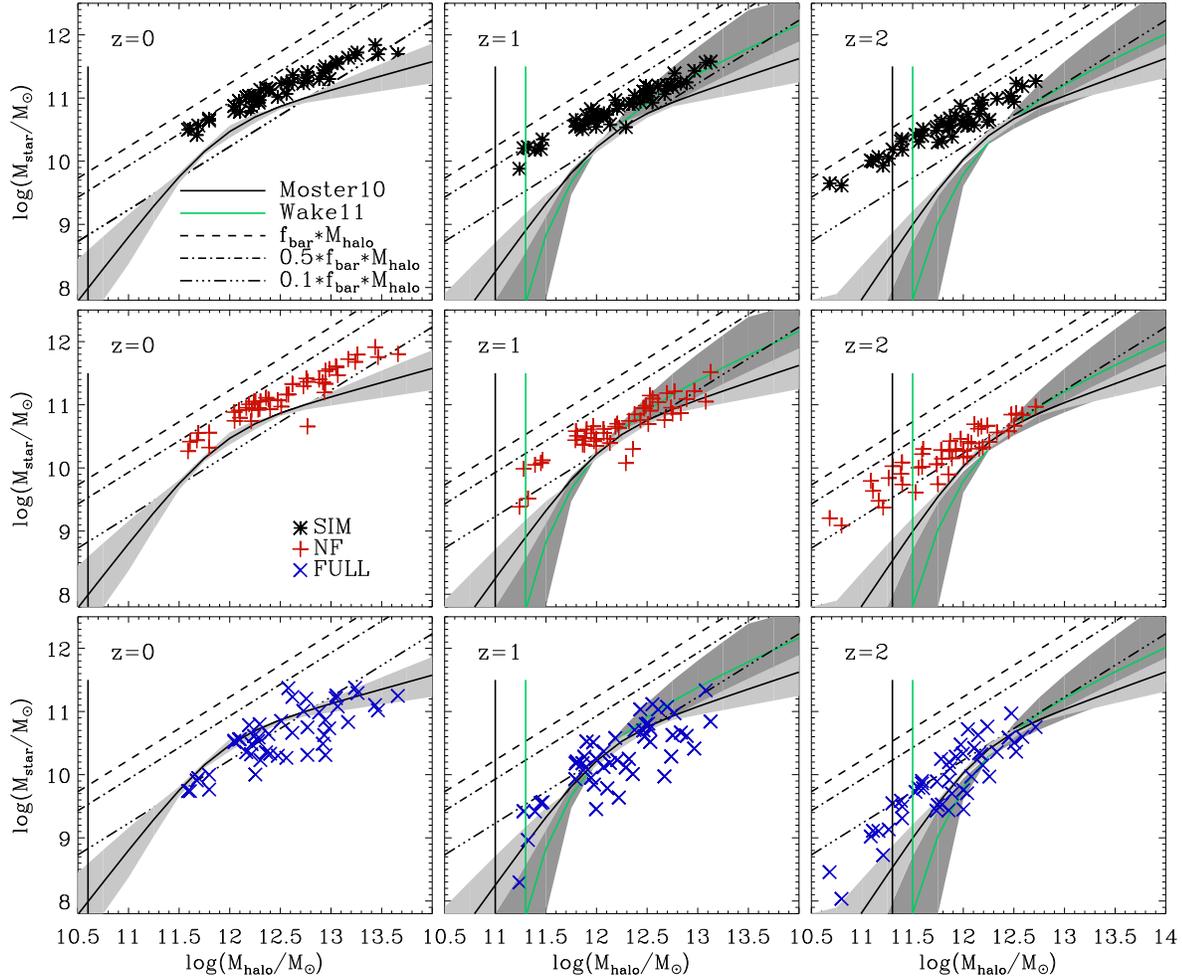, width=0.9\textwidth}
  \caption{Stellar mass versus dark matter halo mass for simulations
    (upper row), NF model (middle row) and the FULL model (lower
    row). The left column shows the dependence on halo mass for
    $z=0$, the middle one for $z=1$ and the right one for $z=2$. The
    black lines with the light grey shaded areas show the fit for the
    halo occupation distribution from \citet{Moster10}. The green
    lines with the dark grey shaded areas the one from
    \citet{Wake11}. The different black, dashed and dashed-dotted
    lines illustrate the total expected baryon fraction
    ($f_{\mathrm{bar}}*M_{\mathrm{halo}}$), 50\% and 10\% of the total
    expected baryon fraction. The black and green thin, vertical lines
    show the observational mass limit of the survey used in
    \citet{Moster10} and \citet{Wake11}, respectively.}  {\label{Msam_Msim}}
\end{center}
\end{figure*}

 This suggests that the agreement presented in \citet{Cattaneo07}
     and \citet{Benson10} is fortuitous, and arises because of the
     enhanced cold flows resulting from the metal cooling included in
     the SAMs. It is unclear whether this agreement would persist if
     metal cooling were also included in the simulations, but it
     appears that the agreement is not nearly as good as they claim
     when metal cooling is omitted from both techniques. Therefore,
     provided that the physical mechanisms taken into account in SAMs
     and simulations are the same, we might conclude that the
     difference between cold and hot mode accretion rates in SAMs and
     simulations is a general result, independent of the specific SAM
     that is considered. This indicates that the
     cooling recipe should be improved in SAMs.
 In addition, the recipes for gas accretion in SAMs do
not currently allow co-existing cold and hot gas accretion as seen in
simulations. For this, \citet{Lu10} proposed a new model that
explicitly incorporates cold-mode accretion independent of the hot
halo gas.  By fitting the hot and cold gas fraction in simulations as
a function of redshift and halo mass and assuming accretion onto the
galaxy within a free-fall time they calculate the accretion rate of
the cold component and thus, achieve a better match of their SAMs to
simulations.

\section{Comparison to observations}\label{comp} 

In this section we compare our results from the SAMs and simulations
to observational data and empirical constraints at different
redshifts. We focus on two key observational constraints: the
relationship between halo mass and stellar mass (the
$M_{\mathrm{gal}}-M_{\mathrm{halo}}$-relation) and the relationship
between stellar mass and star formation rates
($\dot{m}_{\mathrm{star}}-M_{\mathrm{gal}}$-relation).

Fig. \ref{Msam_Msim} shows the relation of galaxy mass and dark matter
halo mass for $z=0$ (left panels), $z=1$ (middle panels), and $z=2$
(right panels). We show the NF and FULL SAMs, and the simulations. We
also show the empirical constraints on the
$M_{\mathrm{gal}}-M_{\mathrm{halo}}$-relation from \citet{Moster10},
which were obtained by asking how halos and sub-halos in a N-body
simulation must be populated in order to reproduce the observed
stellar mass functions at different redshifts (halo abundance
matching). The thin, black vertical lines illustrate the observational
lower observational mass limit. Also shown are the similar
constraints from \citet{Wake11}, which are derived using galaxy
clustering data from the NEWFIRM Medium Band Survey between $1<z<2$
(see also \citealp{Wechsler06, Zheng07, 
  Conroy09, Guo09,Zehavi10, Behroozi10}). Note that the fitting
  functions of \citet{Moster10} and \citet{Wake11} are somewhat
  different, in particular at the low mass end (lower observational
  mass limit is shown by thin, green lines). This is not
  surprising, as they were derived using different methods and from
  different observational data sets.
As our sample consists of only 48 halos covering a mass range
    between approximately $10^{11} M_{\odot}$ and $10^{13} M_{\odot}$,
    the comparison to observational data is not statistically rigorous
    due to the relatively small number of simulated halos. However,
    the 48 halos have been chosen randomly from a cosmological
    simulation and therefore, they should generally follow the mean
    abundance matching trends that relate halo mass to stellar mass.
At all redshifts, the simulations overpredict the stellar masses at a
given halo mass by about a factor of two for halos more massive than
$10^{12}M_{\odot}$. At higher redshifts the progenitor galaxies have
lower masses, and deviate more from the expected distribution. At
$z=2$ the difference can be almost two orders of magnitude for halos
of $\sim 10^{11} M_{\odot}$, in line with the findings of the previous
section --- at high redshift, gas is very efficiently converted into
stars in the simulations. Implementation of more efficient feedback
from supernovae would help to solve this problem.  Indeed it has been
shown that simulations that do include effective SN FB agree much
better with expectations (see
e.g. \citealp{Scannapieco09,Sawala10,Genel10,Governato10} and
references therein).  In addition, it is apparent the simulations
require an additional process that can quench star formation at late
times in massive halos, such as radio mode AGN feedback.

For high mass halos, the NF model (middle row) predicts galaxy masses
close to the relation at $z=1$ and $z=2$ due to less efficient star
formation at high redshifts (see Fig. \ref{mah_stars}), but, like the
simulations, overpredicts the stellar masses in low mass halos. By
$z=0$ the offset is about as large as for the simulations. For the
FULL SAM there is good agreement at $z=0$, which is not surprising
because the model was tuned to match the observed stellar mass
function. However, the FULL model still predicts galaxy masses that
are about a factor of two to three too high for low mass halos
($\log(M_{\mathrm{halo}}) < 11.5$) at high redshift. This is related
to the excess of low mass galaxies at high redshift and other
connected problems discussed in \citet{fontanot:09}, and is seen in
both SAMs and hydro simulations from several groups. It is likely that
these problems are due to limitations in our current understanding or
implementation of the physics of star formation and/or SN feedback. 

\begin{figure}
\begin{center}
  \epsfig{file= 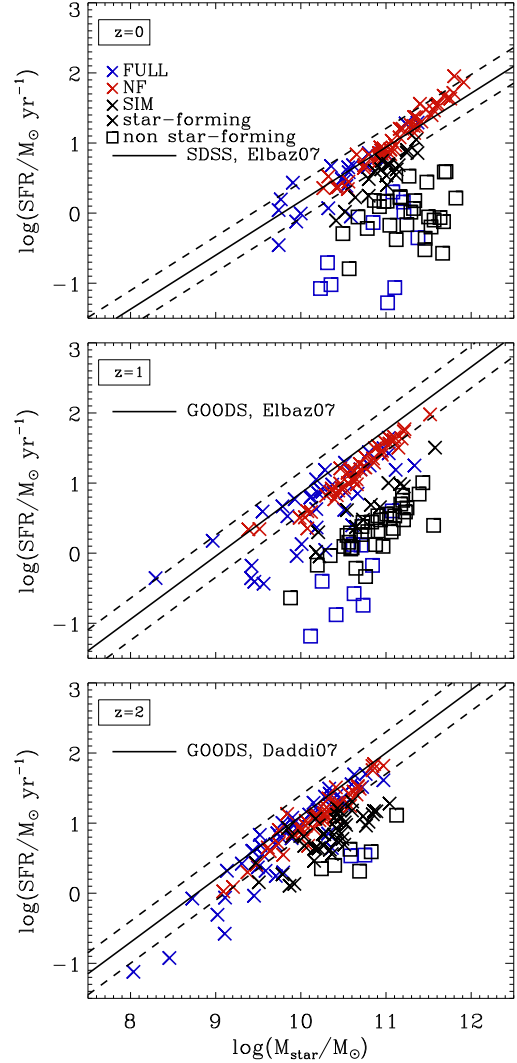, width=0.45\textwidth}
  \caption{Star formation rate versus stellar mass for simulations and
    two different semi-analytic models (NF and FULL). Solid lines
    illustrate the observed relations at different redshifts: $z=0,1$:
    \citet{Elbaz07}, $z=2$: \citet{Daddi07}. Dashed lines illustrate
    the corresponding $1-\sigma$-range of the scatter. Note that in
    light of the recalibration of SFR from 24 $\mu$m observations
    based on Herschel results (see text), we have plotted the Daddi et
    al. relation shifted downwards by 0.3 dex.}  {\label{SFR_Mstar}}
\end{center}
\end{figure}

The relation between the star formation rates and the galaxy stellar
masses $M_{\mathrm{star}}$ is shown in Fig. \ref{SFR_Mstar} for
redshifts $z=0$, $z=1$, and $z=2$. 
We distinguish between star-forming and non-star-forming galaxies
(illustrated as crosses or open squares, respectively) using a
criterion according to \citet{Franx08}: galaxies with specific star
formation rates $SFR/M_{\mathrm{star}}$ smaller than $0.3 \times
t_{\mathrm{hubble}}^{-1}$, are considered to be quiescent, whereas
galaxies with larger specific star formation rates are assumed to be
star-forming. The black, solid lines always refer to the observed
relation at the corresponding redshift ($z=0$: SDSS, \citet{Elbaz07};
$z=1$: GOODS, \citet{Elbaz07}; $z=2$: GOODS, \citet{Daddi07}) for
star-forming galaxies. Note that we show the relation of
\citet{Daddi07} at $z=2$ re-normalised $0.3\ \mathrm{dex}$ downwards,
following the re-calibration of SFR derived from 24 micron luminosity
based on recent Herschel observations \citep{Nordon10}.  We show
    Fig. \ref{SFR_Mstar} in order to illustrate the variation in the
    population of the SFR-$M_{\mathrm{star}}$ plane for different
    implementations of the SAMs and the simulations. Note that this is
    \textit{not} intended to be a quantitative comparison to
    observations, but only a qualitative illustration. In particular,
    we can see that --- with regard to the SAMs --- only in the FULL
    model, do we have SF and non-SF (quenched) galaxies co-existing as
    in the observations. This reflects the well-known need for some
    mechanism, such as AGN feedback, to quench SF in massive
    galaxies.

In general, the simulations underpredict the star formation rates for
star forming galaxies at all redshifts, but most notably at $z > 1$,
due to the high star formation efficiencies at even higher redshifts
and the resulting gas depletion (see section \ref{stellarevol}). Once
again, implementing more effective supernova feedback would presumably
suppress star formation in the small, high redshift progenitors of
these galaxies and result in higher SFR at these redshifts (see
e.g. \citealp{Oppenheimer08,Genel10}). The NF SAMs fit the SF sequence
at $z=0$, but have too many high mass galaxies with high SFR. In the
FULL model, these galaxies are quenched by radio mode feedback, in
agreement with observations (see \citealt{2008MNRAS.391..481S}). Both
the NF and FULL SAMs SF sequences are about a factor of two too low at
$z\sim1$ and 2. This also seems to be generic to many SAMs, as shown
by \citet{fontanot:09} and others. It is interesting to note that star
forming galaxies in both the NF and FULL SAM galaxies all lie on the
same SF sequence. This implies that SN feedback simply moves galaxies
along the sequence, reducing both the stellar mass and the SFR such
that the galaxies remain on the same relation, and is due to the
self-regulating nature of SN feedback in the SAMs. In constrast, AGN
FB quenches star formation and moves massive galaxies off of the SF
sequence (seen in the FULL model).

\section{Discussion and Conclusions}\label{conclusion}

In this paper we present a detailed comparison between a set of $48$
cosmological hydrodynamic zoom simulations and different stripped-down
versions of semi-analytic models based on dark matter merger trees
extracted from the simulations. The hydro simulations are run using
the entropy-conserving formulation of SPH with the \textsc{Gadget}-2
code, and include atomic cooling assuming a primordial composition of
H and He, a UV background radiation field, star formation, and
supernova feedback. We present results from a ``no feedback'' (NF)
version of the SAM which contains the same physical ingredients as the
simulations, and also versions that include thermal feedback by SNae
(SN), a version that also includes large-scale SN-driven winds and
metal cooling (SNWM), and the FULL version which includes all the
previously mentioned ingredients as well as AGN feedback.  With this
approach we can compare the predictions of the two methods over two
orders of magnitude in halo mass at unprecedented resolution.

The two approaches try to answer the same questions but differ in
methodology. In the simulations the full dynamical and hydrodynamical
evolution of the systems is followed by solving the equations of
motion computationally. Additional physical processes like star
formation and supernova feedback are included using sub-resolution
models. The semi-analytic models are based on the computed dark matter
accretion history and approximate the gas physics, star formation, and
feedback processes with simplified recipes. Our analysis is focused on
the cosmic evolution of the baryon content in the central galaxies of
the main branch of the merger trees and its division into various
components (stars, cold gas, and hot gas), as well as how those
galaxies acquired their gas --- whether through ``cold'' or ``hot''
mode accretion --- and their stellar mass (e.g. through in situ star
formation vs. accretion).

The results of our comparison are quite rich, with some surprising
agreement and some striking disagreement. First, we note that we
expect the results of the simulations to lie somewhere in between the
NF and the SN SAMs, since these SAMs include the same physical
processes as the simulations. In most cases, the agreement is best
between the simulations and the NF model, suggesting that the SN FB
implemented in the simulations has little effect. The NF SAMs produce
very good agreement with the simulations for the mass of cold gas plus
stars at all redshifts and for all halo masses. The SAMs slightly
underestimate this ``condensed baryon'' fraction at high redshift
($z>1$) and overestimate it at low redshift. This indicates that the
overall cooling and accretion rates in the SAM and the simulations
must be similar. The NF SAM also produces fairly good agreement
(better than $\sim 20$\% since $z\sim4$) with the overall baryon
fractions (i.e. hot gas plus cold gas and stars in the central galaxy)
in the simulations, here overestimating the baryon fractions at high
redshift in high and intermediate mass halos.

A striking difference is that when we study the evolution of the
stellar and cold gas components separately, we find that the cold gas
fractions agree at very high redshifts, but the gas is consumed much
more rapidly in the simulations, leading to cold gas fractions at all
redshifts less than about $z\sim3$--4, and in halos of all masses,
that are much lower (by up to two orders of magnitude) than in the NF
SAM. Correspondingly, we find much higher stellar masses in the
simulations than the SAMs at high redshift, although they converge to
almost the same value as the NF SAMs at $z=0$. We interpret this as an
indication that the star formation efficiency is much higher in the
simulations than in the SAMs, and the reason for the convergence in
the stellar masses at low redshifts is because nearly all available
gas has been consumed in the simulations. This conjecture is supported
by our finding that if we boost the star formation efficiency in the
NF SAM by a factor of ten, we find excellent agreement with the
stellar mass fraction evolution in the simulations, and improved
agreement with the cold gas fraction evolution.

However, both the simulations and SAMs supposedly adopt the same
empirical Schmidt-Kennicutt relation between cold gas density and star
formation rate. How can the star formation efficiencies be so
different? We note several differences in the implementation of the
star formation recipe in the simulations and SAMs. In SAMs, the only
available information about the structure of the star forming gas in
galactic disks is an estimate of the scale radius of the total
baryonic component of the disk, which comes from angular momentum
conservation arguments \citep{mmw:98,somerville:08b}. The SAMs then
make a series of assumptions --- that the gas is in a smooth, thin
exponential disk with a radius that is a simple multiple of the
stellar scale radius --- and apply the Kennicutt relation in terms of
the predicted gas surface density. Only gas above a critical surface
density is allowed to form stars. In constrast, the simulations
provide detailed 3D predictions for the structure of the cold gas in
galaxies, and implement the SK relation in terms of 3D volume
density (also applying a threshold for SF in terms of a critical
volume density). It is well known that high redshift galaxy assembly
in cosmological simulations is dominated by clumpy, high density, cold
mode accretion. Disks may be more compact than in the idealized case
of perfect conservation of angular momentum, and are thick and
clumpy. The adopted SF recipe is super-linear in the gas volume
density (i.e. the exponent in the SK relation is larger than unity),
and therefore star formation will be more efficient in a clumpy gas
distribution than in a smooth one.

The appropriate values of the SF and SN feedback efficiency parameters
for the simulations were obtained by tuning them to match the observed
Kennicutt relation \emph{for an idealized, smooth thin exponential
  disk}, designed to resemble a Milky Way-like galaxy at $z=0$
\citep{Springel03}. Using these same values for the parameters, we
find that the high redshift galaxies in our cosmological simulations
lie about a factor of five above the Kennicutt relation with the
normalization adopted by \citet{Springel03}. A second, more minor
issue, is that the SAMs were tuned to match a Kennicutt relation with
a normalization a factor of two lower than the one used by
\citet{Springel03}, reflecting a different choice of IMF in the
conversion from observed flux to SFR.

This seems to explain the reason for the factor of $\sim 10$ higher
SFE in the simulations relative to the SAMs, but begs the question: is
this a robust prediction of the simulations that should be taken
seriously? Are these higher star formation efficiencies in high
redshift galaxies really physical? There are several issues that are
relevant here. First, the predicted ``clumpyness'' of the disks is
highly sensitive to the assumed sub-grid recipes. For example,
implementation of more effective SN FB would reduce the clumpyness of 
the disks at high redshift, but might increase the clumpyness at
$z\approx 2$\citep[e.g.][]{Genel10} in the simulations. Observed disks
at high redshift are known to be more clumpy than nearby
ones (\citealp{Genzel06, Genzel08, Genzel10, Foerster09, Foerster11}), 
but it remains highly uncertain 
which recipe for SN FB will produce the ``correct'' degree of
clumpyness and overall structure for statistical samples of high
redshift galaxies while simultaneously reproducing the properties of
local spirals \citep{Piontek11,Governato09, Scannapieco09,
  Brooks09}. Moreover, recent observational studies indicate that star
formation rate densities in local galaxies as well as at high redshift
correlate linearly with the surface density of the \emph{molecular
  gas}, with no evolution in the \emph{molecular} SK relation
\citep{2008AJ....136.2846B,Daddi10,Genzel10}. This is also true for
galaxies with very clumpy star formation, as expected for a linear
dependence with gas density.  Only interacting galaxies undergoing a
significant starburst seem to show an increased star formation
efficiency \citep{Daddi10,Genzel10}. Neither the SAMs nor the
simulations presented here include these effects, although we are
working on updated models that will do so.

A second major difference between the SAMs and the simulations is in
the mode in which galaxies acquire most of their stellar mass. We
distinguish between ``in situ'' growth, due to stars that form out of
cold gas within the galaxy in question, and ``accretion'' of stars
that formed in external galaxies and are accreted via mergers. This
distinction is important because it may determine the characteristic
size evolution of early type galaxies \citep{Khochfar06a, Naab09,
  Guo11, Covington11}. It has been shown previously that massive galaxies in
the simulations and semi-analytic models have an early phase of growth
dominated by the in situ mode, and then switch over to a mode that is
dominated by growth through accretion
\citep{DeLucia06, Khochfar06, Guo08, Oser10, Feldmann10, Zehavi11}. 
Although the ratio of in situ to accreted stars always decreases with
time in the SAMs, in qualitative agreement with the simulations, in
the SAMs in situ growth dominates over accretion in halos of all
masses at all times. Examining the absolute mass in stars formed in
situ or accreted, we find that this is primarily because the SAMs
predict much less mass growth through accretion --- the in situ mass
evolution agrees fairly well with the simulation results. The
comparison between SAMs with different physical processes included
gives us further insights into the origin of the discrepancy: 1)
Increasing the efficiency of SN FB (SN and SNWM SAMs) reduces the
accreted mass because star formation is suppressed in the low-mass
satellites that eventually get accreted. 2) Including radio mode AGN
FB (FULL SAM) reduces the in situ growth in massive galaxies at late
times, because it shuts off the fuel supply for in situ star formation
in these objects.  Interestingly, increasing the SF efficiency in the
SAM does not affect the fraction of in situ versus accreted mass,
presumably because all galaxies are boosted equally.

In summary, it is likely that the contributions from in situ and
accreted stars are currently incorrectly predicted in both the
simulations and SAMs, for the following reasons. In order to match
observations, the simulations presented here clearly require both a
process that suppresses star formation in low mass objects at all
redshifts (such as SN-driven winds) and one that can shut off residual
cooling and quench star formation in massive galaxies at late times
(such as radio mode AGN FB). The former will reduce the accreted mass,
while the latter will decrease the in situ mass in massive objects at
late times. In addition, if the SFE is higher in high redshift
galaxies than at late times, this is likely to increase the accreted
mass. On the other hand, the SAMs presented here, like many SAMs in
the literature, make the assumption that hot gas from the halo can
only be accreted onto the central galaxy. This rapidly truncates the
star formation in satellite galaxies and is likely to artificially
decrease the accreted mass fraction. As a possible solution,
    strangulation of satellite galaxies might
    be delayed if the depth of the potential well of the sub-halo is
    deep enough to retain the hot halo gas for a longer time
    (e.g. Khochfar \& Ostriker 2008). In this case, the
    accreted objects may have longer on-going cooling resulting in
    longer on-going star formation. In addition, the SAMs neglect
gravitational heating, which is clearly important in the simulations
and should reduce the in situ growth in massive halos at late times.
The recent study of \citet{Fontanot11} has shown that the S08 SAM and
other similar SAMs in which radio mode feedback is used to solve the
overcooling problem overpredict the fraction of radio loud galaxies
compared with observations, and the implemented dependence of radio
luminosity on stellar and halo mass is too steep. Gravitational
heating could play a similar role
\citep{Khochfar08,dekel_birnboim:08,birnboim_dekel:10} and thereby
reduce the need for such strong radio mode heating.

A third important result is that the cooling recipe implemented in
these SAMs (which is widely used in many SAMs) overpredicts the
overall accretion rate of gas by a factor of 1.5 (in low mass
halos) to four (in massive halos). If the accretion rates in the
simulations are accurate, this implies that the SAMs that match
present day galaxy properties are compensating for this ``extra''
accretion by artificially ``tuning up'' the feedback. Moreover, when
we divide the gas accretion into ``hot mode'' and ``cold mode''
\citep{Birnboim03,Keres05}, we find that the SAMs systematically
overestimate the hot mode growth (by up to an order of magnitude) and
underestimate the cold mode. As well, in the SAMs without metal
cooling, the cold mode shuts off completely at low redshifts (the
shutoff redshift depends on halo mass), while in the simulations it
declines smoothly but continues to low redshifts. This is likely to be
a result of the fact that, in the SAMs, the criterion for
discriminating between hot and cold mode is based on the assumption of
smooth, spherical halos, while in simulations cold gas can stream into
halos along cold, dense filaments. In addition, in the SAMs, gas is
assumed to accrete either in cold mode or hot mode, but simultaneous
cold and hot mode accretion is not allowed. In the simulations, the
dense, cold streams can penetrate deep into the diffuse hot halos,
allowing for both accretion modes to occur within the same halo
\citep{Keres05,Keres09,2009ApJ...694..396B}. 
We consider this discrepancy to be a general feature of currently
    existing SAMs (see section \ref{hotevol}) and for example, similar
    results were found in the study of Lu et al. (2010). 
    In order to overcome the weakness of SAMs in this respect, they
    proposed a new model that explicitly incorporates cold-mode
    accretion independent of the hot halo gas. By fitting the hot and
    cold gas fraction in simulations as a function of redshift and
    halo mass and assuming accretion onto the galaxy within a
    free-fall time, they calculate the accretion rate of the cold
    component and thus, achieve a better match of their SAMs to
    simulations. However, motivated by hydrodynamical simulations, 
    implementing ’gravitational’ heating might provide a more
    promising mechanism as it treats gas inflow and heating in a
    self-consistent way. Following Khochfar \& Ostriker (2008), the
    heating of the intracluster medium (ICM) would be calculated by
    the net surplus of gravitational potential energy released from
    gas that has been stripped from in-falling
    satellites. Gravitational heating is found to be an efficient
    heating source for massive dark matter halos, where it prevents
    cooling, and becomes especially important at late
    times. Therefore, if we generally assumed cold infalling gas
    and heating of gas due to the binding energy of the halo
    potential, this might imply an automatic, less efficient heating
    at high z and for low mass halos. This should result naturally in
    larger cold accretion fractions for these objects and thus,
    also in a better match of the hot gas content for high-redshift
    galaxies.  

We compared both the simulations and SAMs to two key observational
constraints at $z=0$, 1, and 2: the relationship between dark matter
halo mass and stellar mass, and the relationship between stellar mass
and star formation rate. We found that the simulations predict stellar
masses that are too large for their halo masses at all redshifts. The
stellar masses are too high by a factor of a few for massive halos
($\gtrsim 10^{12} M_{\sun}$), and by an order of magnitude or more for
lower mass halos. The SAM results for the NF model are qualitatively
similar, although the stellar masses at high redshift are lower, due
to the lower star formation efficiencies, as already discussed. In the
SAMs, the curvature in the empirical
$M_{\mathrm{gal}}-M_{\mathrm{halo}}$ relation can be achieved by
including supernova driven winds, which suppress star formation in low
mass halos, and radio mode AGN FB, which suppresses star formation in
high mass halos. We found that the SFR at a given stellar mass were
too low in the simulations at all redshifts $z\lesssim 2$, probably
because of the overly efficient SF at higher redshifts and the
resulting gas consumption. SF galaxies in all SAMs were found to lie
on the same relation, with supernova feedback shifting the galaxies
along the relation. The SAMs (both FULL and NF) showed better
agreement with the observed $\dot{m}_{\mathrm{star}}-M_{\mathrm{gal}}$
relation than the simulations, but have SFR at a given stellar mass
that are about a factor of $\sim 2$ too low at high redshifts.

In final summary, we conclude that on the one hand, we are encouraged
by the robustness of SAMs as a tool for exploring the qualitative
effects of varying the physical ingredients of galaxy formation. On
the other hand, we have also identified several important areas where
the quantitative accuracy of fundamental physical recipes in the SAMs
should be improved, and several physical processes that are missing in
our SAM but which should be included. Additionally, there is a
tendency to treat numerical simulations as ``truth'', but we have
shown that key predictions of these simulations are sensitive to
uncertain sub-grid recipes. We have also highlighted several physical
processes that are neglected in the simulations studied here, but
which appear to be crucial in order to understand the properties of
real galaxies. These include more effective implementation of
supernova-driven winds, chemical enrichment and metal cooling, and a
self-consistent treatment of the growth of and feedback from black
holes. Of course, these are hardly new suggestions, and considerable
progress has been made recently in all of these areas
\citep[e.g.][]{DiMatteo05,Cattaneo05, Sijacki07,
  Oppenheimer08,Booth09, Scannapieco09, Governato10, Schaye10,
  Sawala10, Ostriker10}. In
conclusion, we suggest that using these two complementary techniques
(SAMs and hydrodynamic simulations) together in close coordination may
provide the most powerful approach to understanding galaxy formation
and evolution for the near future.

\section*{Acknowledgments}
This research was supported by the DFG Cluster of Excellence `Origin
and Structure of the Universe'. We thank Simon White for helpful and
interesting discussions. RSS thanks USM for hospitality during
her visit. MH thanks the STScI for hospitality and support for visits. 

\bibliographystyle{mn2e}
\bibliography{Literaturdatenbank}

\label{lastpage}

\begin{appendix}
\section{Effects of numerical resolution}\label{resolution}

\subsection{Dark matter component}\label{darkresol}

\begin{figure}
\begin{center}
  \epsfig{file= 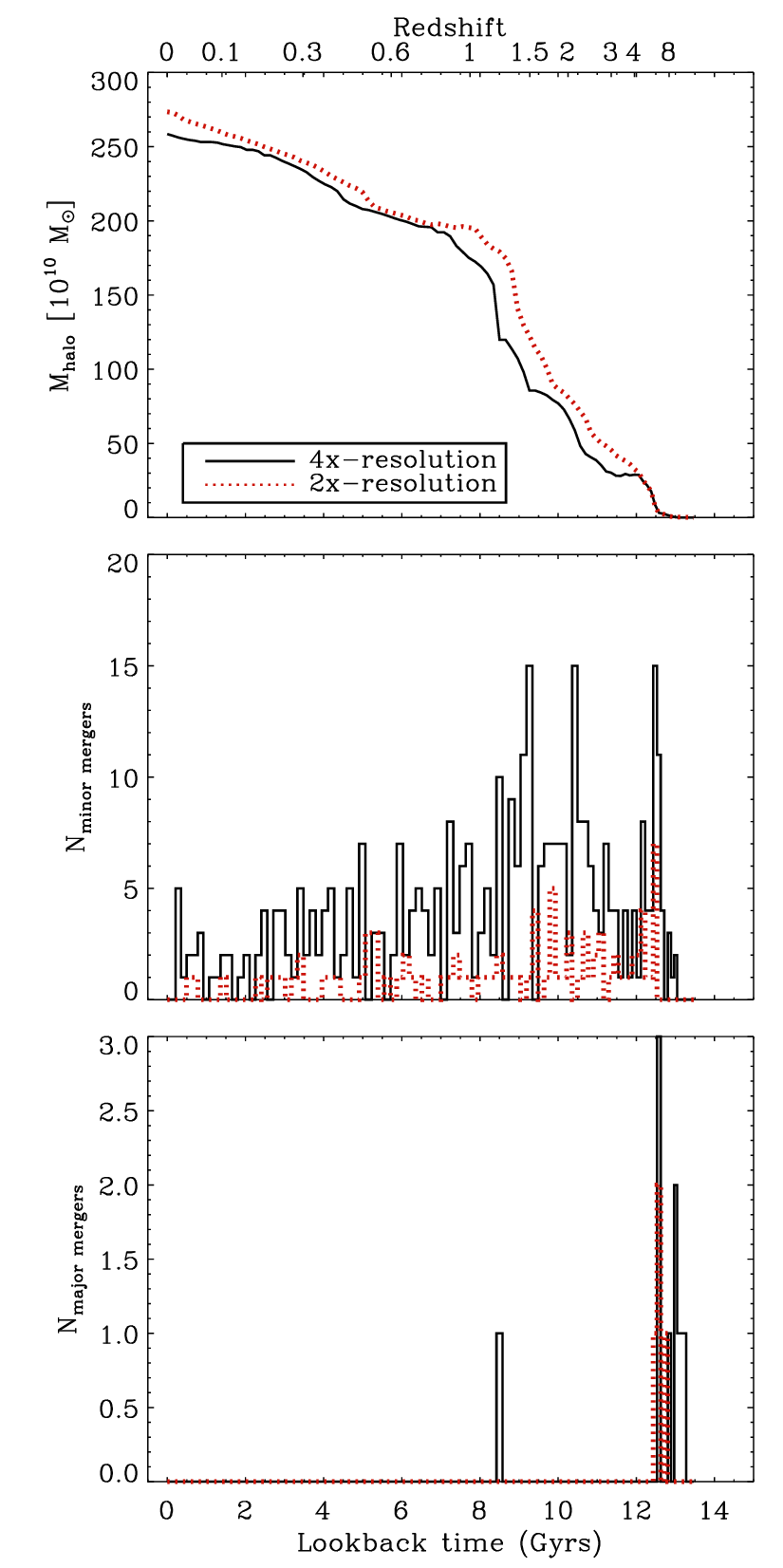, width=0.4\textwidth}
  \caption{Upper panel: comparison of aggregation history for halo
    M0501, with total mass $3.0 \times 10^{12} M_{\odot}$, for two
    different resolutions (2x: red dotted line and 4x: black solid
    line). Intermediate panel: Number of minor mergers ($> 10:1$) as a
    function of lookback time for the two different resolutions. Lower
    panel: Number of major mergers ($< 10:1$) as a function of
    lookback time for the two different resolutions.}
          {\label{mah_resolution_501}}
\end{center}
\end{figure}
\begin{figure}
\begin{center}
  \epsfig{file= 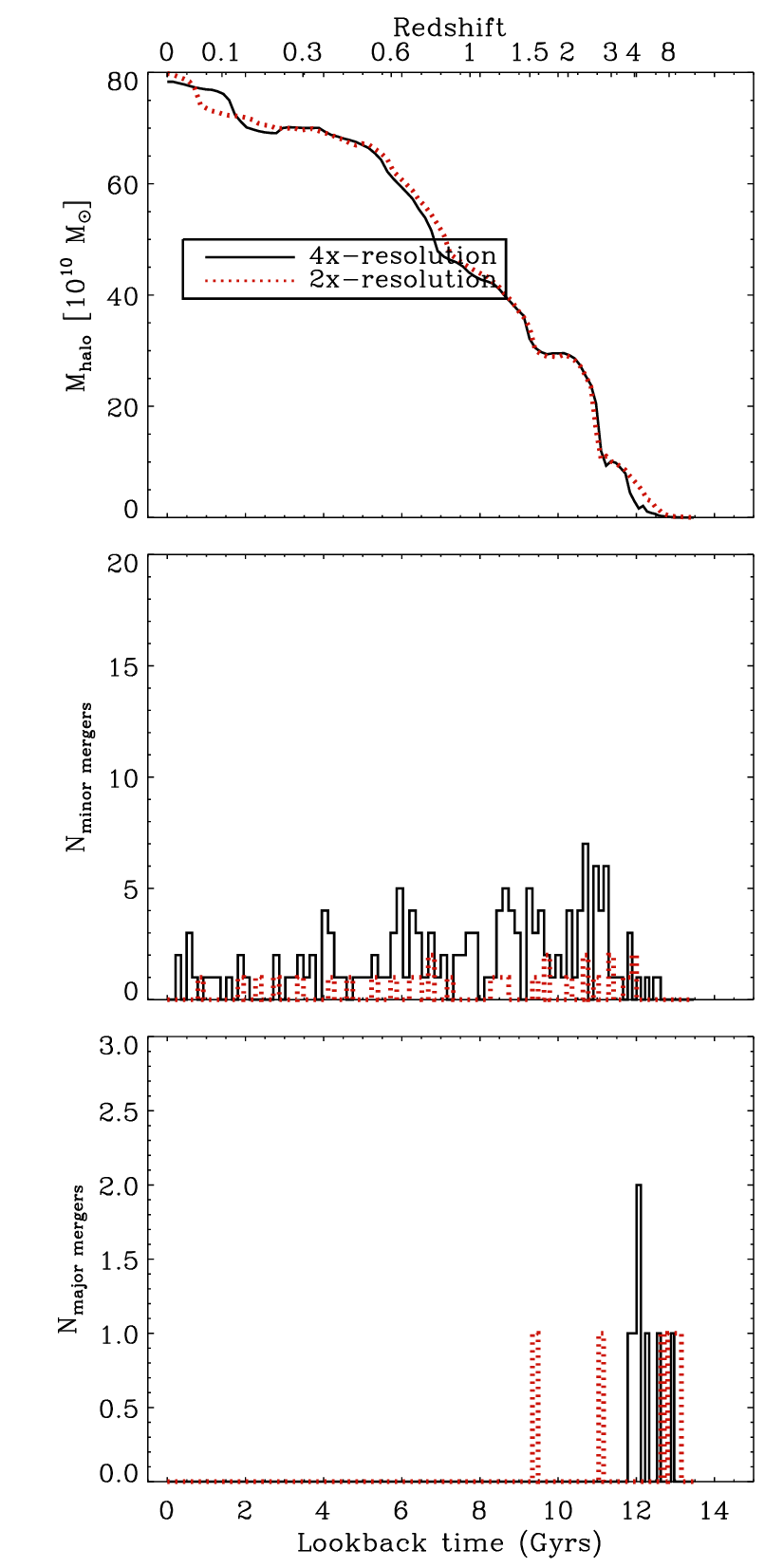, width=0.4\textwidth}
  \caption{Same as Fig. \ref{mah_resolution_501}, but for halo
    M1646, with a smaller mass of $8.0 \times 10^{11} M_{\odot}$.}
          {\label{mah_resolution_1646}}
\end{center}
\end{figure}
\begin{figure}
\begin{center}
  \epsfig{file= 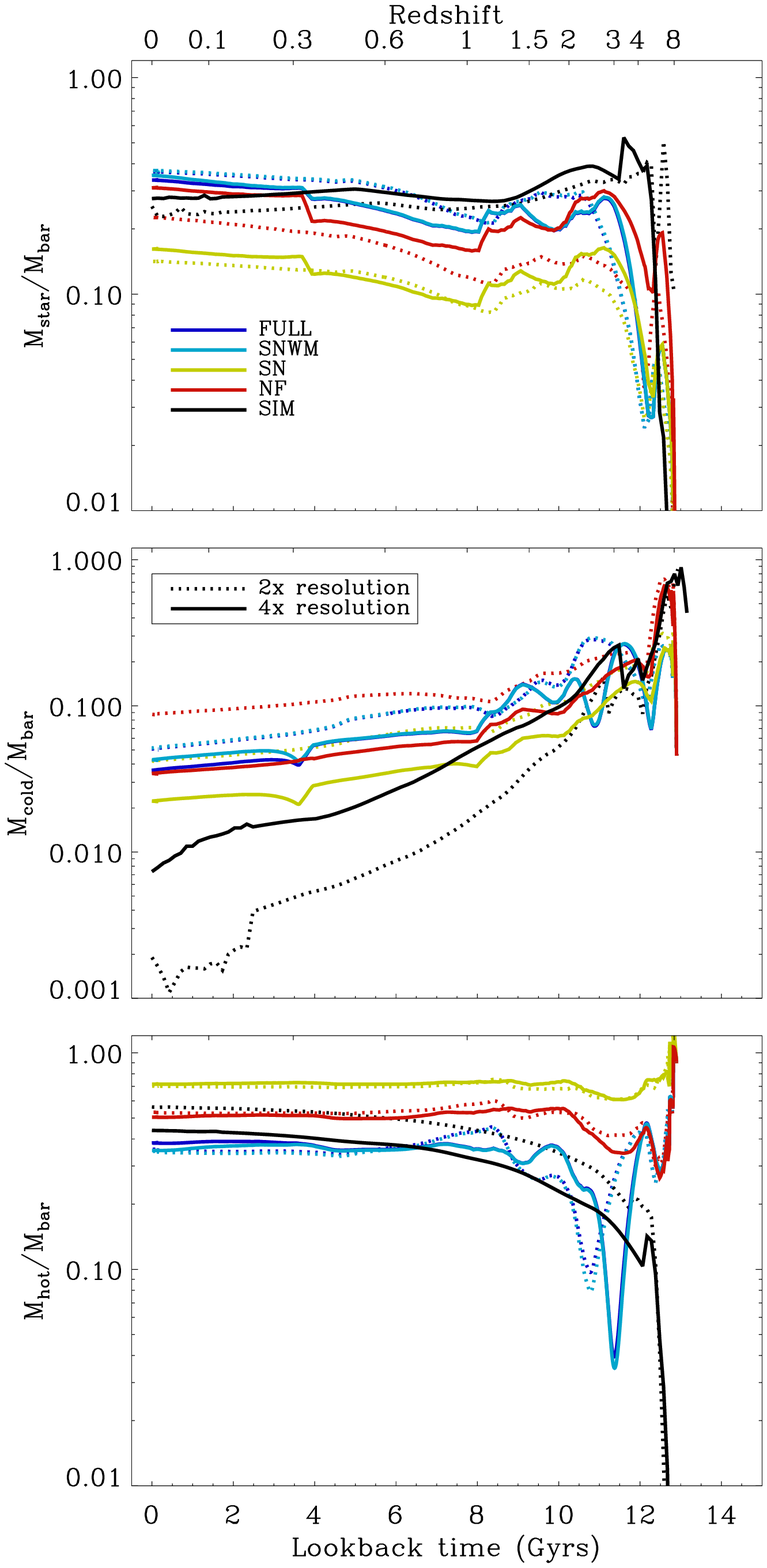, width=0.45\textwidth}
  \caption{Evolution of the baryonic components for two different mass
    resolutions (2x: dotted lines, 4x: solid lines) for halo
    M0501. The upper panel shows the evolution of the stellar
    component, the middle panel the evolution of the cold gas fraction
    and the lower panel shows the hot halo gas component. Black lines
    illustrate the re-simulations, and colored lines show the
    different SAM versions (see main text).}
          {\label{mah_resolution_bar_501}}
\end{center}
\end{figure}
\begin{figure}
\begin{center}
  \epsfig{file= 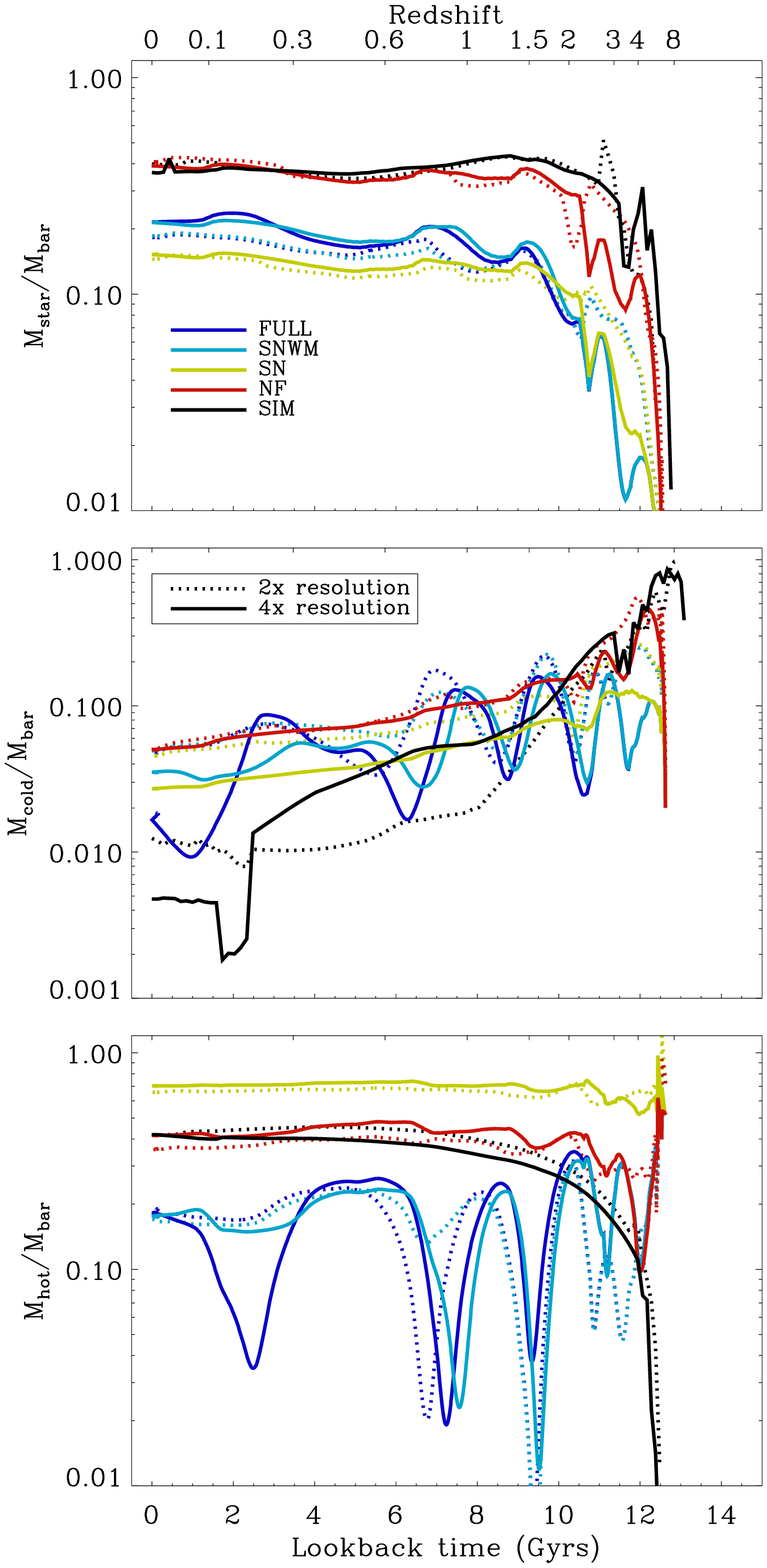, width=0.45\textwidth}
  \caption{Same as Fig. \ref{mah_resolution_bar_501}, but for halo M1646.}
 {\label{mah_resolution_bar_1646}}
\end{center}
\end{figure}
For one high and low mass halo with final masses of $M_{\mathrm{halo}}
= 3 \times 10^{12} M_{\odot}$ (M0501) and of $M_{\mathrm{halo}} = 8
\times 10^{11} M_{\odot}$ (M1646), we have performed re-simulations
with $4 \times$ the original spatial resolution. We traced back the
particles that are closer than $2 \times r_{200}$ to the center of the
halo in any of our snapshots and replace them with dark matter and gas
particles of higher resolution, achieving a $16 \times$ better mass
resolution in the high resolution region than in the original
simulation: $m_{\mathrm{DM}} = 2.5 \times 10^6 M_{\odot}h^{-1}$ and
$m_{\mathrm{Gas}} = m_{\mathrm{Star}} = 5 \times10^5 M_{\odot}h^{-1}$.
The mass aggregation history of the dark matter component in both
halos is very similar for the two different resolution limits as shown
in the upper panel of Figs. \ref{mah_resolution_501} and
\ref{mah_resolution_1646}. 
In the middle and lower panel of Figs. \ref{mah_resolution_501} and
\ref{mah_resolution_1646}, one can see from the histograms of major
($>1:10$) and minor mergers ($<1:10$) that the number of major mergers
stays almost the same, while as expected, we can resolve more minor
mergers ($<1:10$) in the higher resolution case. This suggests that,
since galaxy formation in the SAMs is mainly influenced by the
accretion history of the main halo and by major merger events
($>1:10$), the results from the SAMs are expected to be well-converged
in our re-simulations.

\subsection{Baryonic components}\label{baryonresol}

Figs. \ref{mah_resolution_bar_501} and \ref{mah_resolution_bar_1646}
show explicitly the evolution of the star, cold and hot gas mass in
SAMs and simulations based on the trees from the two re-simulated
individual halos for $2\ \times$ and $4\ \times$ higher resolution. As
expected from the similar evolution of the dark matter component, for
both the low and the high mass halo re-simulations, we find no
significant difference in the evolution of the central galaxy/main
halo in both the SAMs and the simulations.  Therefore, due to
computational costs, we restrict our main study to $2 \times$
re-simulations. 

\end{appendix}

\end{document}